%% file: ms.tex
\documentclass[12pt,preprint,times,longauthor]{aastex63}
\usepackage{epsf,amssymb,amsmath}
\input{defs.tex}
\begin{document}

\title{A Pluto--Charon Concerto II. Formation of a Circumbinary Disk of Debris After the Giant Impact}
\vskip 7ex
\author{Scott J. Kenyon}
\affil{Smithsonian Astrophysical Observatory,
60 Garden Street, Cambridge, MA 02138}
\email{e-mail: skenyon@cfa.harvard.edu}

\author{Benjamin C. Bromley}
\affil{Department of Physics \& Astronomy, University of Utah,
201 JFB, Salt Lake City, UT 84112}
\email{e-mail: bromley@physics.utah.edu}

\begin{abstract}
Using a suite of numerical calculations, we consider the long-term evolution of 
circumbinary debris from the Pluto--Charon giant impact. Initially, these solids 
have large eccentricity and pericenters near Charon's orbit.  On time scales of 
100--1000 yr, dynamical interactions with Pluto and Charon lead to the ejection 
of most solids from the system. As the dynamics moves particles away from the 
barycenter, collisional damping reduces the orbital eccentricity of many particles. 
These solids populate a circumbinary disk in the Pluto--Charon orbital plane; a large 
fraction of this material lies within a `satellite zone' that encompasses the orbits 
of Styx, Nix, Kerberos, and Hydra.  Compared to the narrow rings generated from the 
debris of a collision between a trans-Neptunian object (TNO) and Charon \citep{bk2020}, 
disks produced after the giant impact are much more extended and may be a less promising 
option for producing small circumbinary satellites.

\end{abstract}


\section{INTRODUCTION}
\label{sec: intro}

For nearly a century, the \pc\ system has provided a new window into 
the physical processes that shape planetary systems. As the harbinger 
of the Kuiper belt, the discovery of Pluto contributed a first glimpse 
of the architecture of the trans-Neptunian region. Combined with light
curve data \citep{andersson1973}, the identification of Charon on high 
quality images revealed a tidally-locked binary planet with a short 
orbital period \citep[6.4~d,][]{christy1978}.  Mutual eclipses later 
enabled novel extraction techniques for the characterization of surface 
properties and their evolution with time \citep[e.g.,][]{buie2010a,buie2010b}. 
The discovery of four circumbinary satellites -- Styx, Nix, Kerberos, and 
Hydra -- yielded new insights into circumbinary orbital dynamics and tidal 
evolution \citep{showalter2011,showalter2012,cheng2014b,brozovic2015,
showalter2015}.  Finally, the spectacular results of the \nh\ mission 
illustrated complex geological processes on frozen worlds and made new 
connections between surface phenomena on \pc\ and the physical properties 
of objects in the Kuiper belt \citep[e.g.,][]{stern2015,grundy2016,
mckinnon2016,weaver2016,stern2018,singer2019,mckinnon2020,singer2021}.

Together with the Earth--Moon system, \pc\ establishes interesting constraints 
on the frequency of `giant impacts', collisions between massive protoplanets
often considered as a final stage of planet formation \citep[e.g.,][]{agnor1999,
canup2001,canup2004b,agnor2004,kb2006,asphaug2014}.  In the current picture,
mutual gravitational perturbations place growing protoplanets on crossing orbits 
\citep[see also][]{weth1980,liss1987,weiden1997b,gold2004}.  As protoplanet 
orbits become chaotic, collision velocities become large enough to enable 
catastrophic collisions, which may lead to complete mergers or binary planets, 
with additional ejection of debris \citep[e.g.,][]{canup2005,asphaug2006,
canup2011,genda2012,stewart2012,genda2015a,genda2015b,kb2016a,quintana2016}. 

Although many studies elucidate the physical state of material following 
the Moon-forming impact \citep[e.g.,][]{nakajima2014,pahlevan2016,lock2018,
nakajima2018,tang2020}, there are few investigations of the evolution of 
circumbinary debris following the \pc\ collision. Several analyses explain how 
small satellites on circumbinary orbits become unstable as the central binary 
expands from a period of 1--2~d to the current 6.4~d \citep[e.g.,][]{ward2006,
lith2008a,youdin2012,cheng2014b,bk2015b,woo2018}.  \citet{kb2014b} and 
\citet{walsh2015} consider how collisional damping converts eccentric 
circumbinary orbits into the more circular orbits required for the growth 
of solids into 5--20~km satellites. \citet{bk2015b} demonstrate how a swarm
of cm-sized pebbles could stabilize satellite orbits throughout the tidal
evolution of the central binary.

Here, we examine the evolution of solids orbiting a highly eccentric 
\pc\ binary. Within the \citet{canup2005,canup2011} smooth particle 
hydrodynamics (SPH) calculations, there is a broad range of plausible 
outcomes for the semimajor axis, $a \approx$ 5--30~\rp\ (where \rp\ is 
the radius of Pluto) and eccentricity, $e \approx$ 0.1--0.9, of the
binary planet. The plausible amount of debris surrounding the binary 
is large, ranging from $\sim 10^{20}$~g up to $\sim 3 \times 10^{23}$~g. 
The more likely configurations have a mass in debris that is $\sim$
10--1000 times the total mass of the circumbinary satellites
\citep{kb2019a,kb2019b}.  The goal of the present study is to investigate 
the evolution of solids around very eccentric binaries and to learn whether 
likely end states of the solids in these systems have the properties needed 
to explain the current configuration of the circumbinary satellites in the 
\pc\ system.

The results of this study demonstrate that almost any outcome of a giant 
impact leads to the production of a stable disk of circumbinary solids.  
As the central binary tries to eject circumbinary material close to Charon's
orbit, collisions among the solids rapidly damp eccentric orbits. Among a suite 
of 70 calculations, the amount of solid material in a vertically thin disk 
depends on the properties of the central binary and the initial mass and 
orbital properties of the solids. Wider binaries ($a$ = 10~\rp, $e$ = 0.6--0.8) 
retain a larger fraction of the impact debris than more compact binaries 
($a$ = 5~\rp, $e$ = 0.2--0.4). Circumbinary systems of solids with less mass 
($10^{21}$~g) preserve more of their initial mass than more massive swarms
with $\sim 10^{23}$~g.  All of the 70 calculations maintain enough mass to 
form several 5--20~km satellites.

After establishing the numerical methods for our calculations in 
\S\ref{sec: methods}, we describe several simulations in detail and 
summarize global results in \S\ref{sec: results}. We conclude with a 
discussion (\S\ref{sec: disc}) and a brief summary (\S\ref{sec: summ}).

\vskip 6ex
\section{NUMERICAL METHODS}
\label{sec: methods}

\subsection{Basic Parameters}
\label{sec: pars} 

To set the stage for the calculations, we adopt measured properties of the
\pc\ system \citep{stern2015,nimmo2017,stern2018}. For an adopted gravitational 
constant $G = 6.67408 \times 10^{-8}$, Pluto has mass $m_P = 1.303 \times 10^{25}$~g, 
radius $R_P$ = 1188.3~km, and mean density $\rho_P$ = 1.854~g~cm$^{-3}$.  Charon 
has mass $m_C = 1.586 \times 10^{24}$~g, radius $R_C$ = 606~km, and mean density 
$\rho_C$ = 1.702~g~cm$^{-3}$. Combined with analyses of HST observations 
\citep{brozovic2015,showalter2015}, detailed \nbody\ calculations \citep{kb2019b}
provide limits on the mass of Nix ($m_N \lesssim 4.5 \times 10^{19}$~g) and Hydra
($m_H \lesssim 4.8 \times 10^{19}$~g; see also Youdin, Kratter \& Kenyon 2012). 
Although constraints on the masses of 
Styx ($m_S$) and Kerberos ($m_K$) are more elusive, plausible estimates are 
$m_S \approx 6 \times 10^{17}$~g and $m_K \approx 10^{18}$~g. Thus, the total 
mass in satellites is $m_{sat} \approx 10^{20}$~g. These satellites orbit within 
a `satellite zone,' $a \approx$ 33--66~\rp, defined as a region that encompasses 
the orbits of the four satellites \citep[e.g.,][]{kb2019a,bk2020}.

The current \pc\ system has a semimajor axis, $a \approx$ 16.49~\rp, and an orbital
period $P$ = 6.387~d \citep[see ][and references therein]{stern2018}. The orbit is 
nearly circular \citep{buie2012}; satellite orbits are also nearly circular 
\citep[e.g.,][]{buie2013,brozovic2015,showalter2015}. For the numerical calculations
considered here, we adopt a much more compact binary, $a \approx$ 5--10~\rp, with 
a range of eccentricities, $e \approx$ 0.2--0.8. Tidal evolution models suggest that
these compact binaries evolve into the current architecture on time scales of
$10^6$--$10^7$~yr \citep[e.g.,][]{farinella1979,dobro1997,peale1999,cheng2014a,
correia2020}.  This time scale is much longer than the 1000~yr evolution time 
of a typical numerical calculation for the evolution of circumbinary debris.  
Thus, we ignore tidal evolution.

\subsection{Numerical Code}
\label{sec: code}

To calculate the evolution of circumbinary solids in the \pc\ system, we run numerical 
simulations with \orch, an ensemble of computer codes designed to track the accretion, 
fragmentation, and orbital evolution of solid particles ranging in size from a few 
microns to thousands of km \citep{kenyon2002,bk2006,kb2006,kb2008,bk2011a,bk2013,kb2016a,
knb2016,bk2020}. In this application, we employ several \orch\ tools:
a multiannulus coagulation code to derive the dynamical evolution of small solids and
a gravitational \nbody\ code to enable the solids to respond to the gravitational
potential of the central binary. A set of massless tracers enables communication between
the coagulation and \nbody\ codes.

Using the \orch\ multiannulus coagulation routine, we establish a radial grid, 
centered on the \pc\ barycenter,
of 28 concentric annuli distributed in equal intervals of $a^{1/2}$ between
$a_{in}$ (as listed in Table~\ref{tab: init}) and $a_{max}$ = 160~\rp. 
Each annulus has 80 mass bins with minimum radius $r_{\rm min} = 0.01$~$\mu$m\ and 
maximum radius $r_{\rm max} = 1$~m. In a region between $a_{in}$ and $a_{out}$, we 
seed the grid with solids having total mass $M_0$, material density 1.5~g~cm$^{-3}$, 
surface density $\Sigma \propto a^{-2}$, and a size distribution $n(r) \propto r^{-3.5}$.  
With these initial conditions, most of the mass lies in the largest objects; the
innermost annulus has somewhat more mass than the outermost annulus.  Solids have 
initial pericenter distance $q_0$ and inclination $\imath_0$ as listed in 
Table~\ref{tab: init}.

Within the coagulation code, solids evolve due to collisional damping from 
inelastic collisions and elastic, gravitational interactions.  For inelastic
and elastic collisions, we follow the statistical, Fokker-Planck approaches 
of \citet{oht1992} and \citet{oht2002}, 
which treat pairwise interactions (e.g., dynamical friction and viscous stirring) 
between all objects.  We assume the mass distribution is fixed in time. At the start 
of each calculation, collisions between m-sized (cm-sized) particles are destructive 
(accretive).  Within a day or two, however, collisional damping reduces relative 
velocities by factors of 2--3. Although collisions between m-sized objects remain 
destructive, collisional growth among smaller objects tends to replenish these objects 
about as fast as they are destroyed.  Several tests allowing collisions to grow and 
fragment particles suggests this process modifies the overall size distribution little 
over the course of a typical simulation. 
In these tests, we used the full 
calculational approach for two sets of calculations, where (i) collision outcomes 
are derived and debris is distributed among mass bins and (ii) collision outcomes 
are ignored (maintaining a constant size distribution) but collision rates are used 
to calculate collisional damping as in set (i). Aside from mass redistribution, the 
calculations are otherwise identical, using the elastic formalism to treat gravitational
interactions that do not result in a physical collision and the inelastic formalism
to treat collisional damping.
To save resources for the more cpu-intensive 
\nbody\ calculations, we thus ignore collisional redistribution of mass throughout 
each calculation.

\begin{deluxetable}{lccccccc}
\tablecolumns{8}
\tabletypesize{\normalsize}
\tablenum{1}
\tablecaption{Starting Conditions for Coagulation Calculations\tablenotemark{a}}
\tablehead{
  \colhead{Model} &
  \colhead{$a$ ($R_P$)} &
  \colhead{~~~~~$e$~~~~} &
  \colhead{~~$a_s$ ($R_P$)~~} &
  \colhead{~~$a_{in}$ ($R_P$)~~} &
  \colhead{~~$a_{out}$ ($R_P$)~~} &
  \colhead{~~~~$q_0$~~~~} &
  \colhead{~~$\imath_0$~~}
}
\label{tab: init}
\startdata
1  & ~5.0 & 0.2 & 15.0 & ~8.0 & 32.0 & ~6.0 & ~~0.250 \\
2  & ~5.0 & 0.2 & 15.0 & ~8.0 & 32.0 & ~3.0 & ~~0.250 \\
3  & ~5.0 & 0.4 & 18.3 & ~8.0 & 32.0 & ~7.0 & ~~0.025 \\
4  & ~5.0 & 0.4 & 18.3 & ~8.0 & 32.0 & ~5.0 & ~~0.025 \\
5  & ~5.0 & 0.4 & 18.3 & ~8.0 & 32.0 & ~7.0 & ~~0.250 \\
6  & ~5.0 & 0.4 & 18.3 & ~8.0 & 32.0 & ~5.0 & ~~0.250 \\
7  & ~5.0 & 0.4 & 18.3 & 19.0 & 32.0 & ~7.0 & ~~0.025 \\
8  & ~5.0 & 0.4 & 18.3 & 19.0 & 32.0 & ~5.0 & ~~0.025 \\
9  & 10.0 & 0.6 & 41.5 & 19.0 & 32.0 & 16.0 & ~~0.025 \\
10 & 10.0 & 0.6 & 41.5 & 19.0 & 32.0 & 10.0 & ~~0.025 \\
11 & 10.0 & 0.6 & 41.5 & 19.0 & 32.0 & 16.0 & ~~0.250 \\
12 & 10.0 & 0.6 & 41.5 & 19.0 & 32.0 & 10.0 & ~~0.250 \\
13 & 10.0 & 0.8 & 44.6 & 19.0 & 32.0 & 18.0 & ~~0.250 \\
14 & 10.0 & 0.8 & 44.6 & 19.0 & 32.0 & 10.0 & ~~0.250 \\
\enddata
\tablenotetext{a}{
The columns list the orbital semimajor axis, $a$ in Pluto radii, and eccentricity, $e$, 
of the Pluto--Charon binary; the semimajor axis, $a_s$ in Pluto radii, of the innermost 
stable circular orbit; the initial inner and outer radii, $a_{in}$ and $a_{out}$, and 
the pericenter $q_0$ of circumbinary solids (all in units of \rp); and the initial
inclination $\imath_0$ of circumbinary solids.
}
\end{deluxetable}

To allow the $e$ and $\imath$ of mass bins to react to the gravity of Pluto and Charon, 
we assign massless tracer particles to each mass bin. Among all mass bins, we assign
448,000 tracers.  Within an annulus, each mass bin is allocated $\sim$ 300 tracers 
($\sim$ 24000 tracers for 80 mass bins). Another set of tracers is assigned to bins 
with the most mass per bin.  
The orbital $a$ for each tracer within annulus $i$ is a random number between
$a_i - 0.5 \delta a_i$ and $a_i + 0.5 \delta a_i$, where $a_i$ is the semimajor axis 
of the annulus in the coagulation grid and $\delta a_i$ is the radial width of the 
annulus. With this selection, tracers have semimajor axes that span the full range 
between the inner edge and the outer edge of the coagulation grid. 
Orbital $e_{i,0}$ and $\imath_{i,0}$ for each tracer follow the $e_0$ and 
$\imath_0$ for the assigned mass bin. Tracers with $a$ = $a_i$ have pericenters 
$q_i$ = $q_0$ = $a_i (1 - e_i)$. Other tracers in annulus $i$ have pericenters 
$q_i$ that cluster around $q_0$: 
$q_0 - 0.5 ~ \delta ~ a_i ~ (1 - e_i) \le q_i \le q_0 + 0.5 ~ \delta ~ a_i ~ (1 - e_i)$.  
Because $\delta a_i$ is finite, the tracers have a modest spread in pericenters 
about the adopted $q_0$.
Tracers initially have random orbital phases; some tracers are initially 
near orbital pericenter, while others are near apocenter. Although the midplane 
of the disk lies in the Pluto--Charon orbital plane, some tracers initially lie 
within the orbital plane while others begin their evolution out of the orbital plane.  
Once assigned to a mass bin, tracers may move to another annulus in response to 
gravitational interactions with Pluto and Charon and 
the $(de/dt, d \imath/dt)$ derived in the coagulation code. However,
tracers do not move among mass bins.

Within the coagulation code, each mass bin $k$ in each annulus $i$ contains a 
number of particles $N_{ik}$ and a total mass $M_{ik}$. The average mass of a 
particle is then $ m_{ik} = M_{ik} / N_{ik}$. To allow these particles to 
respond to the motions of the tracers, we assign a number and total mass 
of particles 
to each tracer. With $N_{i,k}^t$ tracers per mass bin, each tracer is responsible 
for $N_{i,k}^{\prime} = N_{i,k} / N_{i,k}^t$ coagulation particles.  To avoid a 
tracer `carrying' fractional coagulation particles, the algorithm assigns integer 
numbers of coagulation particles to each tracer. The mass carried by each tracer 
follows: $M_{ik}^{\prime} = m_{ik} N_{ik}^{\prime}$. Summed over all tracers, the 
mass carried by tracers is the total mass in the coagulation grid.  The number and 
mass of coagulation particles assigned to that tracer effectively follow that tracer 
as it orbits the system barycenter. 

Within the \nbody\ code, tracers are treated as massless particles.  As the 
evolution proceeds, these tracers respond to the gravitational potential of 
the central binary.  In doing so, they generally move to larger distances 
from the barycenter. While some are ejected, others shift from an annulus $i$ 
into another annulus $j$. This shift generates a change in the mass of bins 
in annulus $i$ and a corresponding (opposite) change in the mass of bins in annulus $j$. 
At the end of a time step, the combined set of shifts results in new values 
for the number and total mass of particles in each mass bin. With new numbers 
of tracers in each mass bin, $N_{i,k}^{\prime}$ and $M_{i,k}^{\prime}$ also 
change. Thus, the number of particles and the total mass in those particles 
(but not the average mass of a particle) carried by each tracer varies 
throughout a calculation.

To choose appropriate starting conditions, we examine outcomes of the
\citet{canup2011} SPH calculations. Among this suite of simulations, `successes'
that produce an intact Charon and surrounding debris yield binaries with $a \approx$ 
4--28~\rp\ and $e \approx$ 0.1--0.9. Systems with larger $a$ have larger $e$.  To span 
this range efficiently, we consider binaries with $(a, e)$ = (5~\rp, 0.2), (5~\rp, 0.4), 
(10~\rp, 0.6), and (10~\rp, 0.8). Although these choices do not include the lowest 
$e$ and highest $e$ systems, they are sufficient to learn how the evolution of the 
debris depends on the properties of the underlying binary.

Starting points for the debris surrounding \pc\ follow \citet{canup2011} and models
for the giant impact that produced the Moon \citep[e.g.,][]{ida1997}.
From the \citet{canup2011} calculations, the mass of the debris orbiting \pc\ spans
the range from $\sim 10^{20}$~g to $\sim 3 \times 10^{23}$~g. Considering the initial
orbital state of the debris, we expect considerable losses from dynamical ejections.
Thus, the long-term evolution of a debris field with a mass of $10^{20}$~g is unlikely 
to lead to a stable state with a mass comparable to the current mass of the satellites,
$\sim 10^{20}$~g. We therefore consider masses of $10^{21} - 10^{23}$~g.

To select initial values of $(a, e)$ for debris particles, we match the final maximum
radius of the equivalent circular orbit derived by \citet{canup2011}, 
$a_{eq, max} \lesssim$ 20--30~\rp; \aeq\ is related to the angular momentum of 
orbiting particles \citep[e.g.,][]{canup2004b,canup2011,nakajima2014,bk2020}:
\begin{equation}
\label{eq: aeq}
a_{eq} = a (1 - e^2) ~ .
\end{equation}
Because most of the debris lies near Charon \citep[see Fig. 2 of][]{canup2011},
we consider calculations with pericenters $q_0$ (i) near the apocenter of Charon's 
orbit around the system barycenter or (ii) within Charon's orbit. Assigning a
range of semimajor axes, $a_d \approx$ 8--32~\rp, for the adopted $q_0$ yields 
equivalent circular orbital radii, $a_{eq} \approx$ 7--12, which includes most 
of the outcomes in Fig. 6 of \citet{canup2011}.

Placing the solids within the mass bins of the coagulation grid requires an adopted
surface density distribution $\Sigma(a)$. Independent of the properties of the central 
binary, we adopt $\Sigma \propto a^{-2}$.  Although \citet{canup2011} does not quote 
$\Sigma(a)$ for the debris, \citet{ida1997} consider similar surface density 
distributions in n-body calculations of the formation of the Moon from an extended disk 
surrounding the Earth. SPH calculations for the giant impact do not have the resolution
to derive a size distribution of small particles.  Consistent with the size distributions 
derived for other impact studies \citep[e.g.,][]{green1978,durda1996,durda1997,lein2000,
lein2009}, we adopt a size distribution $n \propto r^{-3.5}$ for the debris 
particles. For simplicity, we assume that all particles within annulus $i$ have the 
same initial eccentricity $e_i$ and inclination $\imath_i$. Thus, particles with 
an initial semimajor axis at the inner edge of the coagulation grid have a smaller
$e_i$ than those with a larger initial semimajor axis.

Compared to the central binary, orbiting solids with the adopted initial surface density 
profile have relatively little total angular momentum.  Following \citet{canup2005}, 
the \pc\ system has an angular momentum
$L_{PC} \approx q \Omega (m_P + m_C) a^2 / (1 + q)^2$, where $q$ is the mass ratio,
$\Omega$ is the angular velocity, and $a$ is the semimajor axis 
\citep[see also][]{canup2011}.  For the adopted masses, 
$L_{PC} \approx 3.5 \times 10^{37}$~g~cm$^2$~s$^{-1}$
($5 \times 10^{37}$~g~cm$^2$~s$^{-1}$) for $a$ = 5~\rp\ (10~\rp). 
The initial configurations of circumbinary solids we consider have 
$L_s \approx 2 - 4 \times 10^{34}$~g~cm$^2$~s$^{-1}$ ($M = 10^{21}$~g) to
$L_s \approx 2 - 4 \times 10^{36}$~g~cm$^2$~s$^{-1}$ ($M = 10^{23}$~g).
For the most massive systems, the angular momentum in orbiting solids is
only $\sim$ 5\% to 10\% of $L_{PC}$.

Table~\ref{tab: init} summarizes the starting points for each calculation.  For each 
combination of $a$ and $e$ of the central binary, we choose two values for the initial 
inclination, 
$\imath_0$ = 0.025 and 0.25. The first choice allows for maximum collisional damping 
throughout the evolution; the second choice provides limits on the ability of collisions 
to evolve a thick disk into the thin disk required for the growth of satellites 
\citep{kb2014b,walsh2015}.  

Among calculations with $e$ = 0.4, we consider two choices for $a_{in}$: 
(i) $a_{in}$ = 8~\rp\ places the solids close to the binary; 
(ii) $a_{in}$ = 18~\rp\ places solids farther out in the system. 
Both of these choices match the $a_{eq, max}$ limits derived in \citet{canup2011}.  
Aside from testing whether the evolution of solids depends on their initial angular 
momentum, these calculations allow us to connect results for compact \pc\ binaries 
with $a$ = 5~\rp\ to those for wide binaries with $a$ = 10~\rp.
In the calculations described below, the mass distribution at 1000 yr depends on
the ability of collisional damping to prevent ejections by the central binary. When
material is close to the binary, collisions and damping are more numerous, but solids
spend more time near the binary. More distant solids have lower particle densities 
and less damping, but these solids spend less time near the binary. Following the
evolution for two sets of starting conditions provides insight into the relative
importance of collisional damping and gravitational perturbations from the binary.

To give the initial conditions of the models additional context, Table~\ref{tab: init} 
lists $a_s$ the semimajor axis of the innermost stable circular orbit for each model. 
For the adopted masses of Pluto and Charon, the numerical results of \citep{holman1999} 
yield
\begin{equation}
\label{eq: a-stable}
\frac{a_s}{a} = 2.175 ~ + ~ 4.58 e ~ - ~ 2.152 e^2 ~ .
\end{equation}
In the compact binaries we consider, $a_s \approx$ 15~\rp\ (18 ~\rp) for $e$ = 0.2 
(0.4). Most of the initial mass then lies outside the innermost stable orbit.  The 
wider binaries have $a_s \approx$ 42~\rp\ (45~\rp) for $e$ = 0.6 ($e$ = 0.8). At the
start of each calculation, most of the mass is on unstable orbits with $a < a_s$.

Our solutions to the evolution equations conserve mass and energy to machine accuracy. 
Timesteps for the coagulation calculations range from a few seconds at $t$ = 0 to 
$10^6$~sec near $t$ = 1000~yr. Within the \nbody\ code, the adaptive integrator divides 
coagulation steps into smaller steps as needed to maintain the integrity of the calculation. 
With 112 cpus, typical calculations require 30000--50000 timesteps and 2--3 cpu sec 
per step. Nearly all of this time is spent evolving the dynamics of tracer particles. 
Over the course of a 1000~yr run, calculations conserve mass and energy to better 
than one part in $10^{10}$.

\subsection{Operations}
\label{sec: ops}

All calculations proceed as follows. For a time step $\Delta t$, the coagulation 
code derives the changes in the $e$ and $\imath$ of each mass bin from collisions 
and orbital interactions with all other mass bins. Collision rates are derived from 
the particle-in-a-box algorithm \citep[e.g.,][]{kb2002a,kb2004a,kb2008}. 
For each mass bin $k$ in annulus $i$, the collision rate with solids 
in mass bin $l$ in annulus $j$ depends on the number density of particles
($n_{ik}$ and $n_{jl}$), the collision cross-section, the relative velocity, 
and the overlap of orbits \citep[see sec.~2 of][]{kb2008}.  When $i = j$,
the overlap is 1; otherwise, the overlap is approximately the ratio of
the volume of annulus $j$ that lies within the volume of annulus $i$.

At the end of the coagulation step, each tracer is assigned a target
eccentricity $e_t$ and inclination $\imath_t$ based on its current
$e_i$ and $\imath_i$ and the changes in $e$ and $\imath$ for its mass
bin from the coagulation calculation. These targets result in time
derivatives for $e$ and $\imath$ during a timestep $\Delta t$, 
$de/dt = (e_t - e_i) / \Delta t$ and
$d\imath/dt = (\imath_t - \imath_i) / \Delta t$.  Before the
\nbody\ step, tracers are also assigned new numbers and masses of
particles based on the number and mass within each mass bin. Although
the \nbody\ code does not use this information, each tracer carries a
changing mass of solids based on the evolution of solids in the
coagulation grid.

Within the \nbody\ step, we solve the equations of motion for a set 
massless tracers orbiting the \pc\ binary, using an adaptive algorithm 
with sixth-order time steps, based on Richardson extrapolation 
\citep{bk2006,bk2011a}.  The code follows objects in the center-of-mass 
frame, calculating forces by direct summation. Aside from the extensive 
tests reported in \citet{bk2006,bk2011a}, the code has been used for 
investigations of migration \citep{bk2011b}, Saturn's rings \citep{bk2013}, 
and the formation and stability of the \pc\ system system \citep{kb2014a,
bk2015a,kb2019a,kb2019b,kb2019c,bk2020}.

Throughout the \nbody\ calculations, tracers evolve with their derived $de/dt$ 
and $d\imath/dt$ and respond to the gravitational potential of Pluto and 
Charon. Orbits evolve with $\sim$ 200 steps per binary orbit.  The tracers' 
eccentricity $e$ and inclination $\imath$ are shifted incrementally at rates set 
by the coagulation code.  These changes to $e$ and $\imath$ are implemented by 
small adjustments to the direction of travel and (if necessary) incremental 
shifts in position, without affecting the other osculating orbital elements. 

At the end of the \nbody\ step, tracers have new positions, velocities, 
and orbital elements $a$, $e$, and $\imath$. Tracers with new $a$ 
that places them outside their 'old' annulus are placed in new annuli;
the number and mass of particles assigned to that tracer move out of 
the old mass bin into a new mass bin within the new annulus. Some tracers
collide with Pluto or Charon; others are ejected beyond the outer
limits of the \nbody\ calculation space, $a \gtrsim 1000~\rp$. After
these tracers are deactivated for the remainder of the calculation,
the coagulation particles associated with these tracers are removed
from the coagulation grid.  When an active tracer has a semimajor axis
outside of the coagulation grid, its mass is removed from the old
annulus and is not placed in a new annulus.  If that tracer returns to
the grid before a collision with Pluto--Charon or ejection, the mass
that it carries also returns to the grid.

Although assigning tracers to annuli based on their current position
$(x, y, z)$ seems reasonable, placement based on $a$ is more in the
spirit of the coagulation code.  The collision and Fokker-Planck
algorithms derive rates based on particle volumes, $V = 4 \pi a \Delta
a H$, where $\Delta a \approx \delta a + e a$, $\delta a$ is the
physical width of the annulus, and $H$ is the vertical scale height
which depends on the orbital inclination $\imath$ \citep{kb2008}. 
For circumbinary solids,
the physical extent of particle orbits is much larger than the physical
width of each annulus. Thus, placing tracers in annuli according to
$a$ recovers the correct volume for calculations of collision and
stirring rates.

New orbital elements for tracers inform the $e$ and $\imath$ of mass 
bins in the coagulation code. Within each mass bin, the median $e$ 
and $\imath$ and their inter-quartile ranges for tracers assigned to 
that mass bin establish the new $e$ and $\imath$ for the mass bin. 
The inter-quartile ranges allow for quantitative monitoring of the 
accuracy of the median in measuring the typical $e$ and $\imath$ 
for a set of tracers.  Typically, the inter-quartile ranges are small.

This sequence provides a closed-loop algorithm that allows the mass 
bins and the tracers to respond to the gravity of Pluto--Charon
and the orbiting solids.  The coagulation particles tell the tracers
how to react to solid material orbiting Pluto--Charon.  In turn, the
tracers tell the solid material how to react to Pluto--Charon. As long
as time steps are not too long, the lag between the coagulation and
\nbody\ steps does not introduce significant offsets in the evolution
of the mass bins and the tracer particles.  

To maximize the accuracy of this approach, the codes have a set of
control parameters. During the coagulation step, changes in the $(e, i)$ 
of particles in each mass bin never exceed 1\%. When changes exceed this
limit, the time step is reduced and repeated. During the \nbody\ step, the 
adaptive integrator adjusts the number of adaptive steps per \pc\ orbit to
track collisions between tracers and Pluto and Charon.
Several examples in 
\citet{bk2006} illustrate the accuracy of the \nbody\ code. After completion
of the \nbody\ step, algorithms within the coagulation code verify that
changes in the median $(e, i)$ among tracers within a mass bin are still 
less than 1\% and that the changes in the inter-quartile ranges among the 
tracers is less than 2\%. Larger changes trigger a reduction in the time step. 
As an independent check on this process, we compared results with the maximum 
level of changes set at 0.5\% and could not measure a significant difference 
between outcomes over 1000~yr of evolution.

\section{RESULTS}
\label{sec: results}

For each of the starting conditions in Table~\ref{tab: init}, we performed simulations 
with five initial masses, $M_0 = 10^{21}$~g, $3 \times 10^{21}$~g, $10^{22}$~g, 
$3 \times 10^{22}$~g, and $10^{23}$~g.  All calculations follow the same evolutionary 
sequence. During their first pass around the barycenter, tracers get a push from 
the central binary. Concurrently, collisions damp the orbital $e$ and $\imath$ 
of every tracer. For the rest of a calculation, the behavior of tracers depends 
on the relative importance of damping and gravitational accelerations that try to remove
solids from the region where circular orbits are unstable. For some tracers, damping 
circularizes their orbits at large $a$ and reduces their inclinations before the binary 
ejects them from the system.  Among other tracers, collisions fail to circularize orbits. 
Although tracers with large initial $\imath$ damp into the orbital midplane of the 
central binary, many are ejected through the midplane beyond the system's Hill sphere. 

After 1000~yr of evolution, most surviving tracers lie within a disk surrounding the 
binary. These tracers have $e \lesssim 10^{-2}$. The disk has a sharp inner edge, where 
solids follow orbits just outside the unstable region, $a \gtrsim a_s$. The peak in 
the disk surface density ($\Sigma$) lies at somewhat larger distances from the binary 
and is within or close to the satellite zone. Beyond the peak surface density at 
$a = a_m$, $\Sigma$ declines rather steeply, $\Sigma \propto a^{-p}$ with $p \approx$ 3--6. 
Among tracers outside of the disk, most have $\imath \approx$ 0; a few have 
$\imath \approx$ 0.05--0.25. Close to the binary (\aeq\ $\approx$ 20--40), 
survivors outside of the disk have $e \approx$ 1.  Tracers with larger \aeq\ have 
a range of $e$; this range grows with increasing \aeq. At \aeq\ $\approx$ 
150--200~\rp, some tracers have $e \approx$ 0.01--0.1.

\begin{figure}[t]
\begin{center}
\includegraphics[width=5.5in]{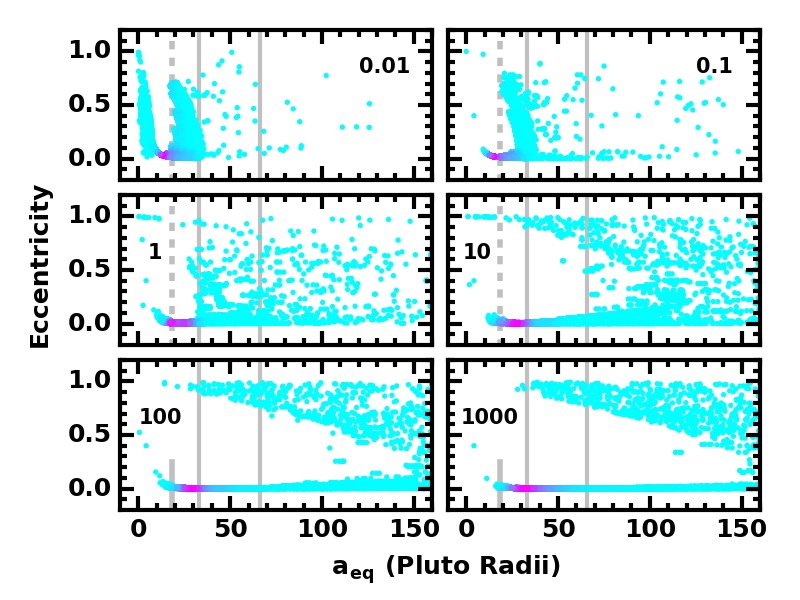}
\vskip -2ex
\caption{\label{fig: ecc1}
Snapshots of the time evolution of material orbiting a central binary with 
$a$ = 5~\rp\ and $e$ = 0.4. Each panel shows the positions of solids in the
$(a_{eq}, e)$ plane for $q_0$ = 7~\rp\ and the times (in yr) indicated in
the legend. Cyan points show positions of individual tracer particles. 
Magenta points denote regions of high density; the intensity of the color 
provides a measure of the density. Initially, most of the solids lie on high
$e$ orbits with pericenter close to the apocenter of Charon's orbit. As the
calculation proceeds, collisional damping places more and more solids into
a disk. Within the disk, the position of peak surface density moves outward
with time.
}
\end{center}
\end{figure}

To illustrate this evolution, Fig.~\ref{fig: ecc1} shows snapshots of the eccentricity 
distribution of tracers for selected times in the evolutionary sequence of model 5 
($a$ = 5~\rp, $e$ = 0.4, $a_{in}$ = 8~\rp, $a_{out}$ = 32~\rp, $q_0$ = 7~\rp, and 
$\imath_0$ = 0.25) 
with an initial mass of $10^{21}$~g.
To place these results in 
the context of previous studies, we plot $e$ (later, $\imath$) as a function of 
\aeq\ instead of the semimajor axis.  Initially, solids have large $e$; $a_{eq}$ 
is then much smaller than the semimajor axis.  As collisional damping reduces $e$, 
$a_{eq} \approx a$.

In this calculation, solids initially have a broad range of $e$, from $e \approx$ 0.1 at 
7~\rp\ to $e \approx$ 0.8 at 32~\rp.  Without collisional damping, all of these orbits
are unstable. However, collisional damping acts rapidly to circularize these orbits.  
After 0.01~yr, some solids lie in a thin ring with \aeq\ $\approx$ 
10--15~\rp\ (upper left panel of Fig.~\ref{fig: ecc1}). Concurrently, the central binary 
attempts to eject particles from the system, placing them within the ensemble of tracers 
at \aeq\ $\approx$ 20--30~\rp\ and $e \lesssim$ 0.6 in this panel. A third set of tracers 
has not had enough time to evolve; these inhabit a group with \aeq\ $\lesssim$ 10~\rp\ and 
$e \lesssim$ 1.

As the evolution proceeds, the tracers with initially small \aeq\ move to larger 
\aeq\ (Fig.~\ref{fig: ecc1}, upper right panel). Some of these lie within the 
narrow ring at small \aeq; others lie within a fledgling disk extending from 
$\sim$ 20~\rp\ to $\sim$ 100~\rp. Almost all of the rest make up the swath of 
particles with a range of $e$ at \aeq\ $\approx$ 20--40~\rp.  However, there are 
a few particles with large \aeq\ and large $e$. These solids are on their way out 
of the \pc\ system.

From 1~yr to 1000~yr, the evolution turns into a competition between collisional 
damping and dynamical excitation by the central binary. Throughout each orbit, 
collisional damping tries to circularize particle orbits. Solids with large $e$ 
receive a kick from the binary near the pericenters of their orbits. This combination 
of processes tends to push particles onto more circular orbits at larger $a$. As 
the system evolves, solids separate into the two groups illustrated in the middle 
and lower panels of Fig.~\ref{fig: ecc1}.  A group of particles with $e \approx$ 0 
forms a thin disk with a peak surface density at 20--40~\rp. This peak gradually 
moves to larger \aeq.  The second group of particles lies within the broad swath at 
the upper right of each panel, with large $e$ at small \aeq\ and small $e$ at large 
\aeq. Within this swarm, collisional damping acts to prevent ejections; combined
with kicks from the binary, particles evolve to larger and large $\aeq$ within the 
disk. For roughly 2/3 of the  particles, collisional damping is too slow; these 
particles are eventually ejected from the system.  Of the remaining solids, nearly 
all lie within the disk. With a final mass of $3 \times 10^{20}$~g, this disk has 
sufficient mass to produce a few larger objects with the combined mass of the current 
satellites ($\sim 10^{20}$~g, Kenyon \& Bromley 2019b).

In the lower right panel of Fig.~\ref{fig: ecc1}, tracer particles outside the
disk constitute only 6\% of the remaining mass in the system. Based on the 
evolution from 100~yr to 1000~yr, we estimate that 1/3 to 1/2 of this material
will be placed into the disk at $a \gtrsim$ 200~\rp. Because these solids will
contribute negligibly to the disk in the satellite zone, we halted the calculation
at $t \approx$ 1000~yr.

\begin{figure}[t]
\begin{center}
\includegraphics[width=5.5in]{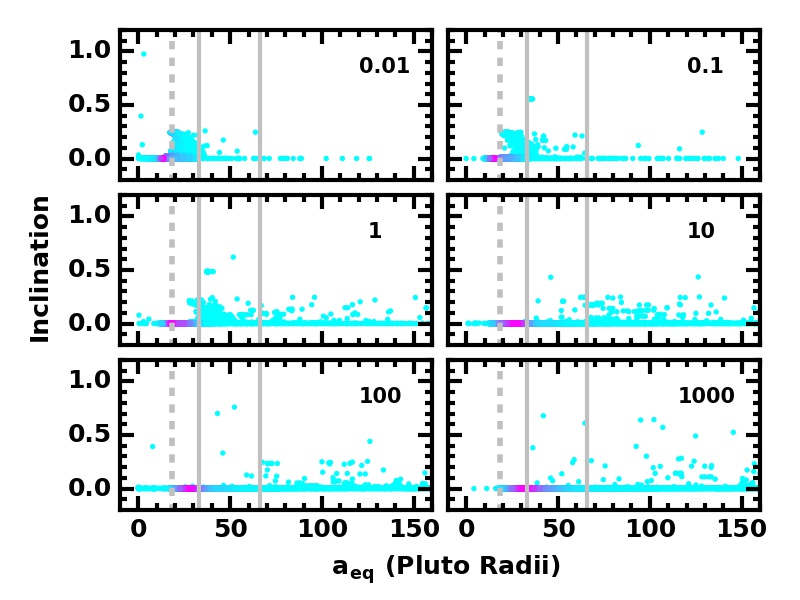}
\vskip -2ex
\caption{\label{fig: inc1}
As in Fig.~\ref{fig: ecc1} for the inclination.  As time proceeds, particles with 
$\imath \approx$ 0.25 move to large \aeq.  Some solids rapidly damp into a thin 
disk with a small scale height. Others are ejected out of the system with a range 
of $\imath$ relative to the orbital plane of the central binary. At the end of the
calculation, most of the remaining solids lie within the vertically thin disk; a 
few with large $\imath$ are on their way out of the system.
}
\end{center}
\end{figure}

The evolution of the inclination closely follows the evolution of the eccentricity
(Fig.~\ref{fig: inc1}). Initially, all tracers have $\imath$ = 0.25. In only a few
days, a substantial fraction of particles has small inclination and lies within the
narrow ring at 10--20~\rp\ or a swarm of high inclination particles with 
\aeq\ $\approx$ 20--30~\rp. Over the next year, many particles damp into the disk
at radii \aeq\ $\approx$ 15--150~\rp. Others remain in a high inclination group that
moves out to 40--50~\rp. Aside from a few outliers, the high inclination group 
disappears from 10~yr to 1000~yr. All other tracers lie within the disk.

At 1000 yr, the disk in Fig.~\ref{fig: ecc1}--\ref{fig: inc1} is well-defined and
fairly stable. Inside 100~\rp, the disk contains a little more than twice the
mass of the \pc\ satellite system, $M_d(a \le 100~\rp) \approx 2.5 \times 10^{20}$~g.
With low eccentricity, $e \lesssim$ 0.01, and lower inclination, $\imath \lesssim$ 0.001,
low mass solids will steadily grow into larger and large objects and eventually form
a robust satellite system \citep{kb2014b}.

\begin{figure}[t]
\begin{center}
\includegraphics[width=5.5in]{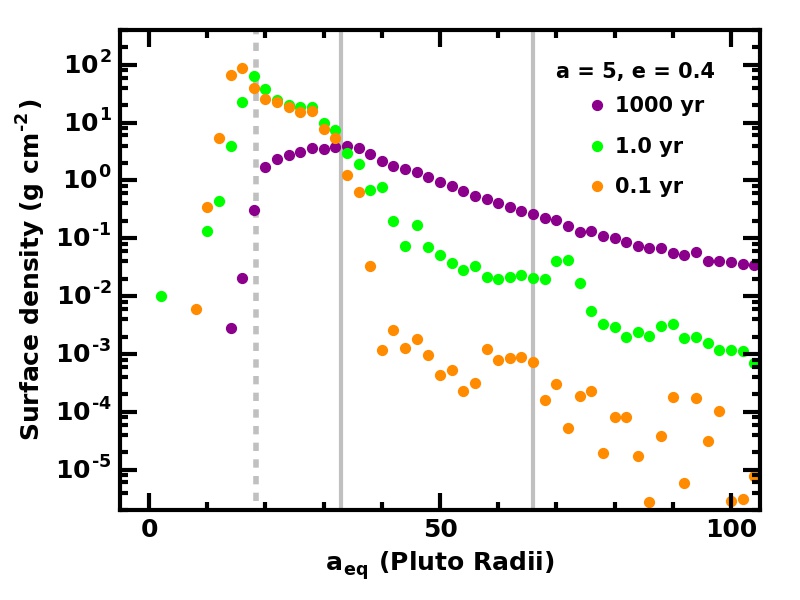}
\vskip -2ex
\caption{\label{fig: sigma0}
Surface density profiles at 0.1~yr (orange points), 1~yr (green points), 
and 1000~yr (purple points) for the calculation shown in 
Figs.~\ref{fig: ecc1}--\ref{fig: inc1}. The vertical dashed line marks the 
position of the innermost stable circular orbit around the central binary.
Vertical solid lines indicate the extent of the satellite zone.
}
\end{center}
\end{figure}

Fig.~\ref{fig: sigma0} shows several snapshots of the surface density profile of 
the disks in Figs.~\ref{fig: ecc1}--\ref{fig: inc1}. After the first year, 
the surface density peaks at 10--20~\rp\ from the system barycenter. For a 
binary with $a$ = 5~\rp\ and $e$ = 0.4, nearly all orbits in this region are 
unstable (Eq.~\ref{eq: a-stable}, Table~\ref{tab: init}). 
As the evolution proceeds, the binary pushes material to larger distances; 
collisional damping reduces orbital $e$ and $\imath$ to keep this material 
in the system.  Over the next 1000~yr, the surface density at 10--20~\rp\ drops 
from close to 100~g~cm$^{-2}$ to less than 0.1~g~cm$^{-2}$.

While the binary ejects solids from the unstable region, damped solids fill the disk
beyond 50~\rp. From 0.1~yr to 1~yr, the surface density at 40--100~\rp\ grows by a 
factor of 100, from $10^{-3}$--$10^{-5}$~g~cm$^{-2}$ to $10^{-1}$--$10^{-3}$~g~cm$^{-2}$.
During the next 1000 yr, the surface density in the disk increases by another factor
of 10. With little change in $\Sigma$ during the last 100--200 years of evolution, the
surface density profile of the disk is stable at the end of this calculation. 

Throughout the evolution, material in the inner part of the disk at 10--20~\rp~ is 
more likely to be ejected than to remain in the disk. At 0.1--1~yr, the inner disk
contains $\sim 10^{21}$~g of solids. After 1000~yr, only $\sim 3 \times 10^{20}$~g
remains at 20--35~\rp. Although some inner disk material finds stability in the 
outer disk at 50--150~\rp, these solids are a small fraction of the material ejected 
from 10--25~\rp. At the end of this calculation, $\sim 10^{20}$~g ($\sim 10^{19}$~g) 
lies at 40--60~\rp\ (60--120~\rp). Overall, the mass of the solids falls from the
initial $10^{21}$~g to a final $3.5 \times 10^{20}$~g. Nearly all of the mass remaining
at 1000~yr is in the disk. 

In addition to a redistribution of mass, the surface density profile gradually 
becomes smoother with time. After a few weeks of evolution, the profile is
irregular and varies chaotically beyond 50~\rp. Within a year, the profile is
smoother, with a wavy appearance beyond 30~\rp. These waves probably result from
the action of the central binary, which perturbs solids as collisional damping
circularizes their orbits. By 1000~yr, the surface density profile is fairly
featureless, with only a few small-scale wiggles superimposed on a monotonically
decreasing profile.

The model 6 calculations follow nearly the same evolution shown in 
Figs.~\ref{fig: ecc1}--\ref{fig: sigma0}. With a smaller $q_0$, solids 
have a larger initial $e$ but the same initial $\imath$. As the binary 
attempts to kick solids out of the system, collisional damping tries to circularize their orbits. Solids that
cross Charon's orbit feel larger gravitational perturbations than those that always
stay outside of the binary orbit. Larger kicks make it harder for damping to reduce 
$e$ and $\imath$ and to keep these particles in the system. Despite the actions of
the central binary, damping reduces inclinations fairly rapidly, generating a thin
disk in the \pc\ orbital plane within 1~yr. With more material in the midplane at 
all times, damping rates grow, enabling a larger fraction of existing particles 
to remain in the system.

Compared to model 5, these calculations exhibit two interesting features. For the
low mass calculations ($M_0 = 1 - 3 \times 10^{21}$~g), collisional damping tends
to lag gravitational perturbations from the binary. Although $e$ and $\imath$ drop
with time, perturbations from the binary push particles to larger and larger $a$.
At larger $a$, the solids occupy a somewhat smaller volume due to the smaller $e$
(which limits their radial excursions) and the smaller $\imath$ (which limits their
vertical excursions), which increase rates and hence the effectiveness of damping. 
Combined with smaller perturbations from the binary at large $a$, enhanced damping
allows formation of a disk at $a \gtrsim$ 30--40~\rp.  After 1000~yr, remaining 
solids have formed a disk with a size similar to the solids in model 5. However, 
these disks have less mass than the model 5 disks. Their shallower surface density 
distributions have a larger fraction of mass at larger distances, a characteristic 
of the initially larger kicks from the central binary and the ineffectiveness of
collision damping at small semimajor axis, $a lesssim a_s$.

In contrast, calculations with the model 6 initial conditions and large initial
masses are able to retain a larger fraction of their initial mass. When the initial 
mass is larger, collisional damping -- which scales as the square of the mass -- is 
more effective. The vertical extent of the swarm decays more rapidly into the 
midplane; more particles lie within the dense ring at 20--30~\rp\ and within the 
nascent disk at larger distances. As the calculations proceed, the binary still 
manages to eject most of the initial mass from the system.  However, the final disk 
masses are remarkably similar to those in model 5. Although there are modest 
differences among the calculations, massive disks with the model 5 starting 
conditions have disks with a physical structure close to the structure of disks 
with the model 6 parameters (see Table~\ref{tab: results}).

\begin{figure}[t]
\begin{center}
\includegraphics[width=5.5in]{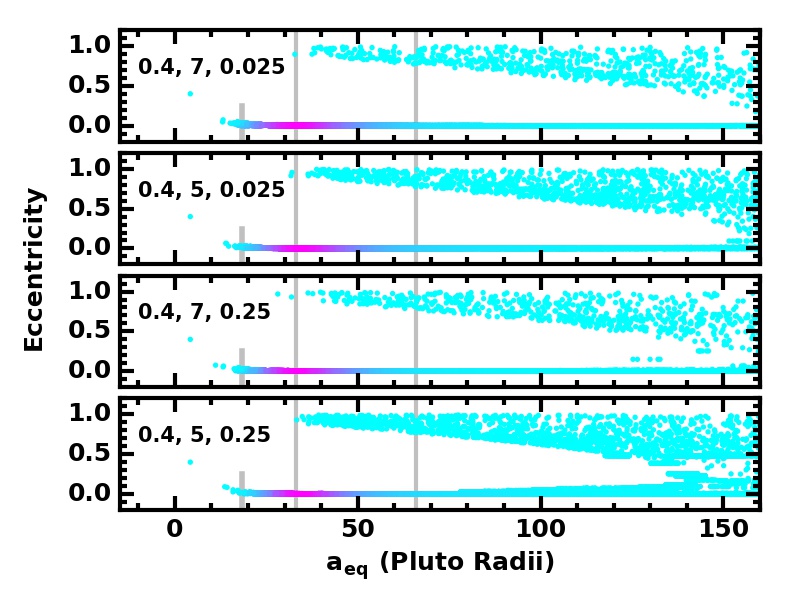}
\vskip -2ex
\caption{\label{fig: ecc2}
Distribution of surviving particles in $(\aeq, e)$ space at 1000~yr for calculations 
with the model 3--6 initial conditions and \M0\ = $10^{21}$~g.  Cyan points indicate 
the positions of individual tracers. Magenta pixels denote regions of high density; 
the intensity of the color provides a measure of the particle density in each pixel.  
Labels list $e$ for the central binary, $q_0$, and $\imath_0$.  In all panels, the 
cyan and magenta points delineate an extended disk with an inner edge at $\sim$ 20~\rp, 
a maximum density at 30--40~\rp, and an outer edge at 200--300~\rp.  The morphology 
of the points in the upper and the lower middle panels for $q_0$ = 7 demonstrates that
the outcome is fairly independent of $\imath_0$. Comparison of the upper two and lower 
two panels illustrate the impact of $q_0$. When $\imath_0$ = 0.25, the outcome at 
1000~yr is fairly independent of $q_0$.  At 1000~yr, systems with larger $\imath_0$ 
have more particles at higher eccentricity than those with smaller $\imath_0$.  
Above the disk, particles in limbo have $e \approx$ 1 at small \aeq\ and 
a broad range of $e$ at large \aeq. Over time, some of these particles will be 
incorporated into the disk; the central binary will eject others from the system.
}
\end{center}
\end{figure}

Reducing the initial inclination of the particles ($\imath_0$ = 0.025, models 3--4)
has a modest impact on the results. With a smaller inclination, the solids occupy
a smaller volume. Collisional damping is then more effective. At the same time, 
particles on eccentric orbits see larger perturbations in the gravitational potential
due to the central binary. As particles experience larger accelerations from the
binary, collisional damping circularizes their orbits somewhat more rapidly compared
to solids in the model 5--6 calculations. Overall, these two processes approximately
cancel: the model 3 and 4 calculations retain roughly the same amount of mass as
their model 5 and 6 counterparts; the overall extent and mass distribution within
the disks are also fairly similar.

Fig.~\ref{fig: ecc2} compares the distribution of $e$ as a function of \aeq\ for 
surviving particles at 1000~yr in calculations with the model 3--6 initial conditions 
and \M0\ = $10^{21}$~g. The upper two panels show results for models 3 and 4; the lower 
two panels plot outcomes for models 5 and 6.  All four panels have a similar morphology. 
A large swarm of particles lies in the orbital plane of the binary. The inner edge 
of this distribution has \aeq\ $\approx$ 20~\rp; the outer edge is beyond the right 
limits of the plot.  Each swarm has a density maximum at 30--40~\rp.  The position 
of the density maximum varies little from one panel to the next; there is no obvious 
dependence of the disk morphology on $q_0$ or $e_0$. 

\begin{figure}[t]
\begin{center}
\includegraphics[width=5.5in]{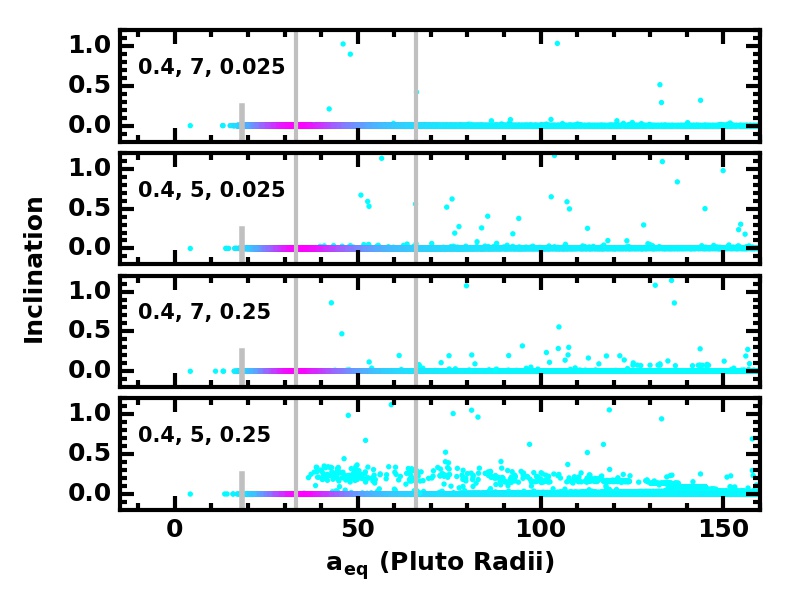}
\vskip -2ex
\caption{\label{fig: inc2}
As in Fig.~\ref{fig: ecc2} for the inclination. At 1000 yr, all calculations show a 
clear disk in the midplane of the \pc\ binary; the disk begins at an inner edge of 
20~\rp, reaches peak surface density at 30--40~\rp, and extends to 200--300~\rp beyond 
the right edge of the panels.  Above the disk, particles in limbo have $\imath \approx$ 0.25 
at small \aeq\ and smaller $\imath$ at large \aeq. Over time, some of these particles 
will be incorporated into the disk; the central binary will eject others from the system.
}
\end{center}
\end{figure}

In all four panels, there is a second swath of particles extending from 
$(\aeq, e)$ = (30, 1) to larger \aeq. At large \aeq, the particles have 
a broad range of eccentricity, $e \approx$ 0.2--1.  This subset of particles 
is in limbo: collisional damping has kept them within the \pc\ system, but 
continued kicks from the binary have prevented them from becoming incorporated 
into the disk. Panels for $q_0$ = 7~\rp\ clearly have fewer particles in this 
swarm than those with $q_0$ = 5~\rp. At \aeq\ $\approx$ 140--160~\rp, there is 
also a clear gap between the disk particles and those outside the disk. In the 
panels for $q_0$ = 5~\rp, this gap is occupied by other particles in limbo. 

Based on the evolution up to 1000~yr, we anticipate that many of the particles
in limbo will eventually become part of the disk. In all of these calculations,
the lower edge of the swarm in $(\aeq, e)$ space forms a nearly straight line from
$(\aeq, e) \approx$ (35, 1) to $(\aeq, e) \approx$ (140--180, 0.1--0.2). Surviving
solids above this locus will have final disk radii beyond 140--180~\rp\ and 
contribute little mass to the disk. 

Fig.~\ref{fig: inc2} repeats Fig.~\ref{fig: ecc2} for the inclination. As in 
Fig.~\ref{fig: ecc2}, all of the panels display a similar morphology.  A large 
swarm of particles with $\imath \approx$ 0 extends from $\sim$ 20~\rp\ to beyond 
150~\rp. With $\imath \approx$ 0, these disk particles lie in the \pc\ orbital 
plane.  Peak surface density within this disk is at 30--40~\rp.  Unlike 
Fig.~\ref{fig: ecc2}, the upper three panels in Fig.~\ref{fig: inc2} show few 
particles outside of the \pc\ orbital plane . Although some of these survivors 
might eventually become part of the disk, the binary will probably eject most 
of them.

The lowest panel in Fig.~\ref{fig: inc2} has a large group of particles from 
$(\aeq, \imath)$ = (40, 0.25) to $(\aeq, \imath)$ = (150, 0.). These particles 
are also in limbo; the competition between collisional damping and excitation 
from the binary is more balanced than for the stable particles already in the 
midplane and for the less fortunate solids ejected from the system. From an 
analysis of the evolution at 10--1000~yr, particles within this swath will 
eventually land in the disk at large \aeq\ or will be ejected from the 
\pc\ system.  The trend in $\imath$ with \aeq\ indicates how evolution into 
the disk proceeds: as damping reduces $\imath$, the binary pushes a particle 
to larger \aeq. The location where the swath crosses the midplane is the 
location where particles damp to roughly zero inclination. As the evolution 
proceeds beyond 1000~yr, we expect that some of particles with higher 
inclination will damp into the disk at \aeq\ larger than 150~\rp. Because our 
main interest is on the location of solids near the satellite zone, we chose 
not to follow this evolution in detail.

\begin{figure}[t]
\begin{center}
\includegraphics[width=5.5in]{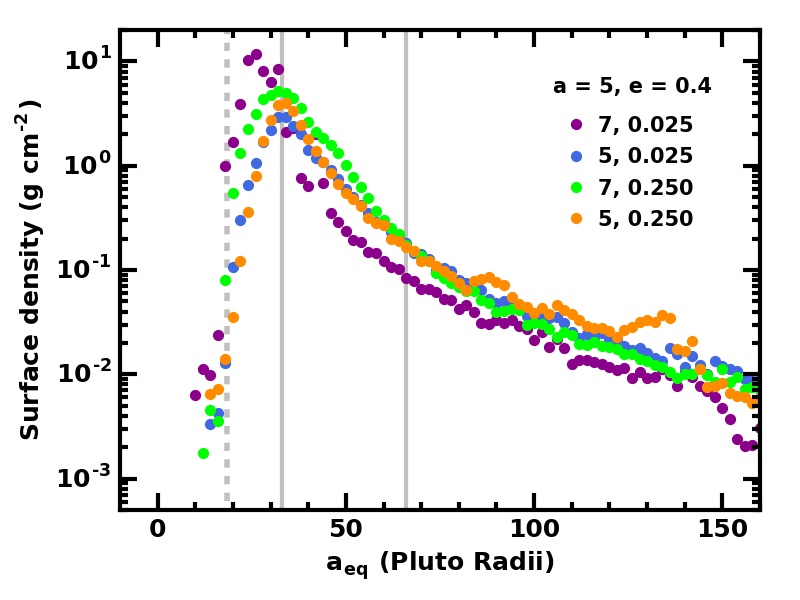}
\vskip -2ex
\caption{\label{fig: sigma1}
Surface density distributions for model 3--6 calculations with \M0\ = 
$10^{21}$~g. The legend indicates $(a, e)$ for the central binary and 
$(q_0, \imath_0)$ for the solids in each calculation. Vertical grey lines
show the boundaries of the satellite zone. The surface density rises from 
a sharp inner edge at $\sim$ 20~\rp\ to a sharp peak at 35~\rp; the surface 
density than falls over two orders of magnitude from the peak to 150~\rp. 
Ripples in the surface density at 80--150~\rp\ for the model with 
$(q_0, \imath_0)$ = (7, 0.25) are typical of disks that have not quite 
settled into an equilibrium state.
}
\end{center}
\end{figure}

Fig.~\ref{fig: sigma1} compares the surface density distributions for models
3--6 and \M0\ = $10^{21}$~g. All exhibit the same general trend: a sharp edge
at 20~\rp, a clear maximum at 30--35~\rp, and a gradual decline out past 
150~\rp. Aside from this trend, there are several clear differences. Models
with $q_0$ = 7~\rp\ ($q_0$ = 5~\rp) have a broad (narrow) peak. The shape
and location of the peak does not depend on $\imath_0$. At large \aeq, the
decline of the surface density is fairly independent of $q_0$ and $\imath$;
in all calculations $\Sigma$ declines by a factor of 300--500 from 
30--35~\rp\ to 150~\rp.

Although three of the surface density distributions have a monotonic decline
from 35~\rp\ to 150~\rp, results for model 6 show a clear wave at 80--150~\rp.
In this example, the surface density profile evolves as shown in
Fig.~\ref{fig: sigma0}. After one year, disk material lies close to the 
innermost stable orbit. As the system evolves, most of these solids are 
ejected. However, some solids in the inner disk move onto stable orbits well 
outside the innermost stable orbit. As the outer disk stabilizes, the surface 
density has a wavy appearance generated by the interplay between the binary 
trying to excite the orbits and collisional damping working to maintain solids 
on circular orbits. During these interactions, solids find a stable $a$ and $e$.
The time scale for the waves to damp (and for the surface 
density to become smooth) depends on the properties of the inner binary and
the initial $q$ of circumbinary solids. In less eccentric binaries, the waves 
damp more rapidly compared to the waves in disks surrounding more eccentric
binaries. Similarly, waves generated by solids with initially smaller $q$ 
take longer to damp than those with initially larger $q$.

In the example shown in Fig.~\ref{fig: sigma1}, the wave results from particles 
with initially large $q$. At later times, the particles have modest eccentricity, 
$e \lesssim$ 0.1, and negligible inclination, $\imath \lesssim 10^{-2}$; 
collisional damping has not quite circularized their orbits. These particles 
are visible as a thickening of the $e$ vs \aeq\ distribution at large \aeq\ in 
the lowest panel of Fig.~\ref{fig: ecc2}. In the other calculations shown in 
Fig.~\ref{fig: ecc2}, collisional damping operates somewhat faster and 
circularizes nearly all of the orbits in 1000~yr.  In model 6, another 1000 yr
of evolution would allow collisional damping to finish circularizing the orbits 
of these particles; the surface density distribution will then follow the other 
results more closely.

Although mean-motion resonances (MMRs) may eventually appear in these surface 
density distributions, none are present after 1000 yr of evolution. For the
calculations illustrated in Fig.~\ref{fig: sigma1}, low order MMRs such as
the 3:1 to 7:1 lie inside the innermost stable orbit for an $e$ = 0.4 binary.
In these four examples, the 14:1, 15:1, 16:1, and 17:1 MMRs at 29--33~\rp\ are
within the region of peak surface density at $\sim$ 30~\rp. However, the time 
scale for MMRs to develop in this region is $\gtrsim$ 0.1~Myr \citep[see, 
for example][]{kb2019a}. As the binary expands on similar time scales 
\citep[e.g.,][]{farinella1979,dobro1997,peale1999,cheng2014a,correia2020}, 
lower order MMRs will sweep through this region \citep{ward2006,lith2008a,
cheng2014b}. Collisional damping can stabilize solids against the perturbations 
induced by these MMRs \citep{bk2015b}.

\begin{figure}[t]
\begin{center}
\includegraphics[width=5.5in]{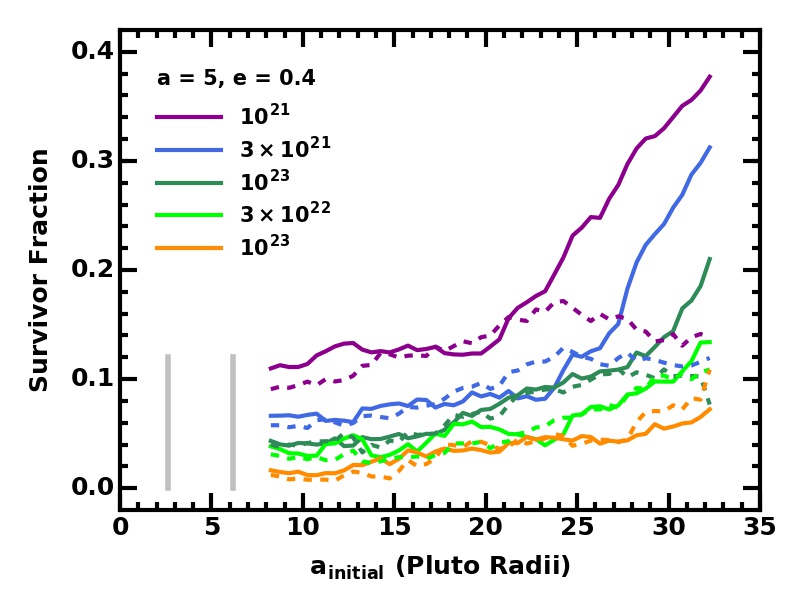}
\vskip -2ex
\caption{\label{fig: survivors}
Fraction of tracers that survive 1000 yr of evolution as a function of their
initial semimajor axis for a binary with $a$ = 5~\rp, $e$ = 0.4, and various 
initial masses as indicated in the legend. Solid (dashed) curves show results 
for systems of solids with pericenters near the apocenter of Charon's orbit 
(with pericenters midway between the apocenter and pericenter of Charon's
orbit), indicated by the vertical grey lines). Although the ability of the 
central binary to solids at 8--25~\rp\ is fairly independent of \qz, the 
retention of solids on more distant, more eccentric orbits depends on \qz.
}
\end{center}
\end{figure}

To illustrate the impact of the initial $a_i$ and \qz\ on the evolution of solids,
Fig.~\ref{fig: survivors} plots the "survivor fraction" as a function of the initial
semimajor axis of solids. For solids with initial semimajor axis at 8--10~\rp, from
2\% ($\M0 = 10^{23}$~g) to 10\% ($\M0 = 10^{21}$~g) remains in the system after 1000~yr.
These fractions are amazingly independent of the initial $q$.  At 10--20~\rp, the 
survivor fraction remain fairly independent of initial $a$ and $q$ for a given 
initial mass. At 20--25~\rp, however, systems with larger initial $q$ retain a
larger and larger fraction of solids as the initial $a$ increases. In contrast,
systems with an initial $q$ at 5~\rp\ have a low survivor fraction
independent of initial $a$. 
Clearly, systems with larger $q$ retain much more of their initial
mass than those with smaller $q$.

The variation of survivor fraction with initial $a$ and $q$ is a result of the
interplay between gravitational forcing from the binary and collisional damping. 
In the restricted three-body problem, perturbations in the potential due to the
binary are a strong function of the semimajor axis, with leading terms 
$\propto a^{-2}$ inside Charon's orbit and $\propto a^{-3}$ outside the binary
\citep{lee2006,leung2013,bk2015a,bk2021}. 
Near the innermost stable orbit, the binary pumps up the eccentricity of
circumbinary solids, or ejects then directly during strong, close
encounters. For a \pc\ binary with $a$ = 5~\rp\ and $e$ = 0.4, the
innermost stable orbit is at $a$ = 18~\rp; inside this distance, objects 
on completely circular orbits cannot survive. Solids on eccentric orbits
are rapidly ejected. At larger distances, near the peak of the surface
density distribution ($a \approx$ 35~\rp), damping is fast enough to
preserve solids even when their initial eccentricity is high.

Collisional damping depends on the particle density $n$ and the relative 
velocity $v$ as $n^2 v$. Initially, particles have large $e$ and occupy a
large volume; collisional damping rates are then low. The binary proceeds
to eject particles by increasing their $a$ and $e$. As $a$ grows, however, 
particles spend less time near the binary. At large $a$, collisional 
damping dominates perturbations from the central binary. Damping lifts the 
pericenters of particle orbits, reducing binary perturbations near pericenter.  
As orbits circularize, $n^2$ grows faster than $v$ falls; damping rates grow 
rapidly. Throughout circularization, particles orbit at larger $a$ due to 
binary forcing, but with smaller $e$ and larger $q$.  Although moving to 
larger $a$ reduces perturbations from the binary {\it and} damping rates, 
damping tends to dominate when particles have larger $a$ and smaller $e$. 
Thus, orbits circularize at semimajor axes much larger than the original 
semimajor axis.

Although damping is effective at large $a$ when solids have pericenters 
near Charon's apocenter (e.g., $q \approx Q_c$), the binary dominates when 
$q \ll Q_c$ \citep[e.g.,][]{lee2006,leung2013,bk2015a,bk2021}. Inside 
Charon's orbit, most orbits are chaotic except for a few stable regions 
near Pluto \citep{winter2010,giuliatti2013,giuliatti2014,giuliatti2015}.
When solids with $a \gtrsim$ 18~\rp\ travel inside Charon's orbit, 
strong, repeated interactions with the binary carry a high risk of
collision or ejections. For some tracers, damping raises the pericenter 
sufficiently to escape these outcomes. For most tracers, however, close
encounters dominate; the binary ejects the tracer from the system before 
damping acts. In this way, solids with initial $q \ll Q_c$ have a small 
survivor ratio independent of initial $a$. 

Tracer survival also depends on the initial mass of the swarm. Because damping
depends on $N_{ik}^2$, systems with more mass damp more effectively. With all 
orbits passing close to (or within) the binary, very efficient damping has the 
drawback of circularizing many orbits at semimajor axes $a \lesssim a_s$. These
orbits are unstable; thus, these particles are ejected rapidly. More massive
systems of solids damp a larger fraction of their mass inside the region where
orbits are unstable. In this way, more massive systems lose a larger fraction 
of their initial mass to dynamical ejections.

Calculations with the model 7--8 parameters begin with the same binary parameters
and the same initial masses and inclinations as models 3--4, but have a different 
radial distribution of solids. With an inner radius of 19~\rp\ instead of 8~\rp, 
all of the solids begin with large eccentricities, $e \approx$ 0.63--0.75 for 
model 7 and $e \approx$ 0.74--0.84 for model 8. Models with larger initial $e$ 
allow the binary more chances to eject particles before collisions circularize 
orbits. Compared to the results for model 5 (model 6), the model 7 (model 8)
calculations retain less mass in the circumbinary disk. Aside from the lower mass,
the disks have roughly the same extent and radial distribution of mass in all of
the models (see Table~\ref{tab: results}).

Despite differences in the amount of retained mass, all of the model 7--8 
calculations evolve in a similar way. Solids near the center-of-mass receive an
immediate kick from the binary. For solids farther away, collisional damping has
a few days to reduce $e$ and $\imath$ before the gravity of the binary impacts 
their trajectories. After the first few orbits, collisional damping eases some
solids into a circumbinary ring at 20--30~\rp\ and a few others into a disk at 
30--100~\rp. As the ring and disk form, the binary ejects many other solids from
the system. Within the first year or two, the ring and disk are fairly well-defined;
as times goes by, the ring expands and damping places more solids into the disk.

\begin{figure}[t]
\begin{center}
\includegraphics[width=5.5in]{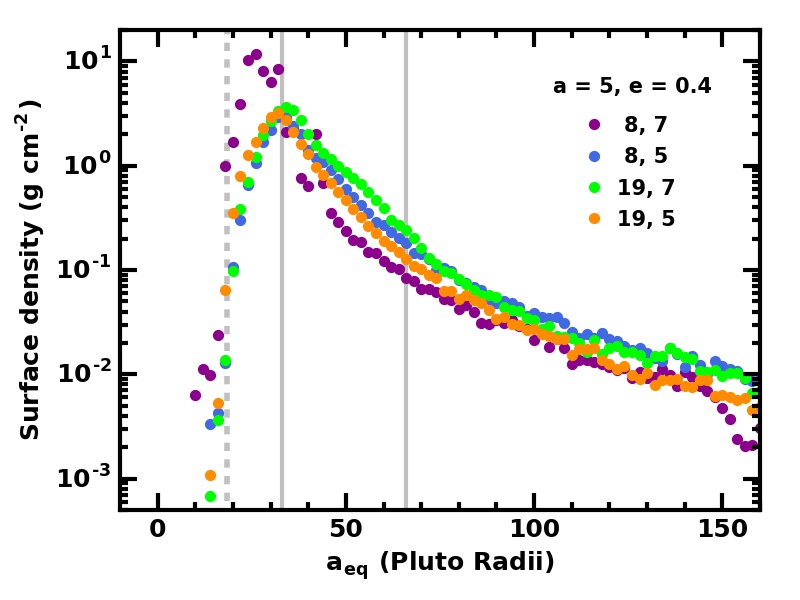}
\vskip -2ex
\caption{\label{fig: sigma2}
Surface density distributions for model 5--8 calculations with \M0\ = 
$10^{21}$~g. The legend indicates $(a, e)$ for the central binary and 
$(q_0, \imath_0)$ for the solids in each calculation. The vertical grey
lines show the boundaries of the satellite zone.  The surface density 
rises from a sharp inner edge at $\sim$ 20~\rp\ to a sharp peak at 35~\rp; 
the surface density than falls over two orders of magnitude from the peak 
to 150~\rp. Ripples in the surface density at 80--150~\rp\ for the model
with $(q_0, \imath_0)$ = (7, 0.25) are typical of disks that have not
quite settled into an equilibrium state.
}
\end{center}
\end{figure}

Fig.~\ref{fig: sigma2} compares the remarkably similar surface density 
distributions at $t$ = 1000~yr for models 3, 4, 7, and 8 and $M_0 = 10^{21}$~g. 
All have negligible mass inside 20~\rp, a sharp peak at 30--40~\rp, and a 
steep decline in $\Sigma$ from the peak to 200--300~\rp. Distributions for
models 7 and 8 are somewhat wavier at 100--200~\rp\ than for models 3 and 4.
Another 1000~yr of damping would smooth out any waves in all of the surface
density distributions. Despite the wave, it is clear that these systems have
much of their mass inside the satellite zone.

Calculations with larger initial mass in solids follow the same evolution
path as those with the lowest mass. When \M0\ is larger, collisional damping 
is more effective; orbits circularize more rapidly. Because the solids evolve
towards orbits with smaller $a$ and shorter periods, the binary has more 
opportunities to eject solids from the system. Over the first few years of 
evolution, systems with larger \M0\ suffer more ejections than those with 
smaller \M0. Survivors are placed on orbits with larger $a$.

\begin{figure}[t]
\begin{center}
\includegraphics[width=5.5in]{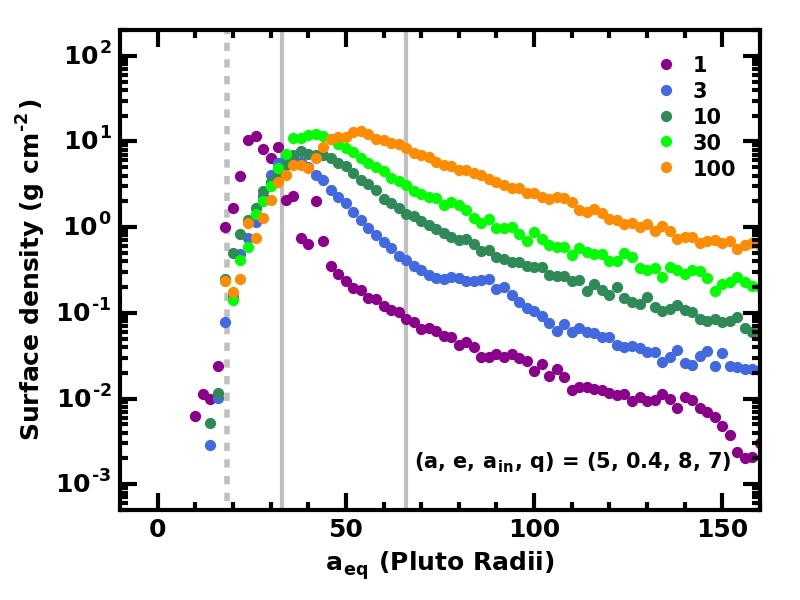}
\vskip -2ex
\caption{\label{fig: sigma3}
As in Fig.~\ref{fig: sigma2} for model 5 and various \M0\ as indicated in the 
legend. Systems with larger initial mass have surface density maxima at larger
semimajor axes.
}
\end{center}
\end{figure}

Fig.~\ref{fig: sigma3} shows density distributions at 1000~yr for model 5 
calculations with different \M0. At \aeq\ = 10--25~\rp, the five calculations 
have a sharp edge, where $\Sigma$ grows by 3--4 orders of magnitude. Each disk 
has a clear maximum in $\Sigma$ at 30--60~\rp; the position of the peak moves
to larger \aeq\ when \M0\ is larger. Beyond the peak, $\Sigma$ falls rapidly
with \aeq. For several calculations, the drops in $\sigma$ is somewhat wavy.
Over time, these waves decline in amplitude.

In these calculations, the slope of the surface density distribution is a strong
function of the initial mass. In systems with the smallest \M0\ ($10^{21}$~g),
$\Sigma$ roughly follows a power-law with slope $p \approx$ 3.5. As the initial mass 
grows, the slope of the power-law drops; for \M0\ = $10^{23}$~g, $p \approx$ 2.5.
The shallower slope is a signature of the processes that generate the disk in 
the plane of the binary's orbit. In more massive systems, the binary is more 
effective at placing material at larger distances, reducing the slope in $\Sigma(a)$
relative to systems with less initial mass.

Despite the sharp peaks in $\Sigma$, surviving solids beyond the satellite zone
contain a large fraction of the final mass. In the most massive models with
\M0\ = $10^{23}$~g, roughly equal amounts of mass lie within the satellite zone
and at \aeq\ $\gtrsim$ 75~\rp. In model 5, for example, the mass in the satellite
zone is roughly 100 times the mass of the \pc\ satellites. In the least massive
models, the sharper drop in $\Sigma(\aeq)$ leaves a smaller fraction of the final
mass in solids beyond the satellite zone. 

Outcomes of calculations with different combinations of $(a, e)$ for the 
\pc\ binary and $q_0$ for the solids lead to similar results. For each model
and \M0, Table~\ref{tab: results} summarizes the mass remaining in active
tracers at 1000~yr ($M_f$), the mass in disk material with $e \le$ 0.1 ($M_d$), 
the inner and outer edge of the disk ($a_1$, $a_2$), the location of peak 
surface density ($a_m$), and the maximum surface density ($\Sigma_m$). Disk 
edges are defined as the point where the surface density first rises above 
$10^{-2}$~\gcms\ (inner edge) or where the surface density first drops below 
$10^{-2}$~\gcms\ outside the peak surface density.

\begin{deluxetable}{ccccccccccccccccc}
\tablecolumns{7}
\tabletypesize{\scriptsize}
\tablenum{2}
\tablecaption{Disk Properties at 1000 yr\tablenotemark{a}}
\tablehead{
  \colhead{Model} &
  \colhead{~~$M_0$~~} &
  \colhead{~~$M_f$~~} &
  \colhead{~~$M_d$~~} &
  \colhead{~~$a_1$~~} &
  \colhead{~~$a_2$~~} &
  \colhead{~~$a_m$~~} &
  \colhead{~~$\Sigma_m$~~} & &
  \colhead{Model} &
  \colhead{~~$M_0$~~} &
  \colhead{~~$M_f$~~} &
  \colhead{~~$M_d$~~} &
  \colhead{~~$a_1$~~} &
  \colhead{~~$a_2$~~} &
  \colhead{~~$a_m$~~} &
  \colhead{~~$\Sigma_m$~~}
}
\label{tab: results}
\startdata
~1 & ~~1 & 0.28 & 0.26 & 16 & 150 & 30 & ~6.39 && ~8 & ~~1 & 0.19 & 0.18 & 18 & 134 & 32 & ~3.21 \\
~1 & ~~3 & 0.79 & 0.74 & 16 & 166 & 28 & 33.69 && ~8 & ~~3 & 0.48 & 0.46 & 16 & 172 & 32 & ~4.34 \\
~1 & ~10 & 1.17 & 1.03 & 12 & 194 & 32 & 19.33 && ~8 & ~10 & 0.89 & 0.82 & 14 & 216 & 40 & ~4.63 \\
~1 & ~30 & 1.97 & 1.59 & 14 & 276 & 44 & 10.43 && ~8 & ~30 & 2.15 & 1.94 & 18 & 278 & 52 & ~6.42 \\
~1 & 100 & 5.48 & 4.52 & 24 & 360 & 44 & 26.31 && ~8 & 100 & 6.25 & 5.19 & 14 & 330 & 56 & 16.44 \\
~2 & ~~1 & 0.11 & 0.09 & 16 & 124 & 26 & ~3.25 && ~9 & ~~1 & 0.68 & 0.67 & 36 & 176 & 52 & ~5.35 \\
~2 & ~~3 & 0.32 & 0.28 & 14 & 166 & 32 & ~4.05 && ~9 & ~~3 & 1.47 & 1.43 & 36 & 180 & 70 & ~7.81 \\
~2 & ~10 & 0.65 & 0.52 & 16 & 168 & 32 & ~6.96 && ~9 & ~10 & 2.66 & 2.47 & 36 & 200 & 70 & ~9.71 \\
~2 & ~30 & 1.65 & 1.32 & 16 & 216 & 42 & ~6.46 && ~9 & ~30 & 4.98 & 4.37 & 34 & 320 & 72 & 14.04 \\
~2 & 100 & 4.10 & 3.23 & 24 & 328 & 44 & 17.49 && ~9 & 100 & 7.78 & 6.51 & 34 & 304 & 70 & 26.63 \\
~3 & ~~1 & 0.14 & 0.13 & 16 & 122 & 32 & ~2.11 && 10 & ~~1 & 0.68 & 0.67 & 36 & 176 & 64 & ~6.63 \\
~3 & ~~3 & 0.56 & 0.50 & 16 & 170 & 36 & ~6.61 && 10 & ~~3 & 1.12 & 1.07 & 36 & 180 & 52 & ~7.41 \\
~3 & ~10 & 1.17 & 1.00 & 16 & 196 & 38 & ~7.72 && 10 & ~10 & 2.08 & 1.91 & 34 & 202 & 52 & ~9.78 \\
~3 & ~30 & 2.49 & 1.86 & 20 & 266 & 42 & 12.18 && 10 & ~30 & 4.40 & 3.85 & 34 & 272 & 72 & 13.39 \\
~3 & 100 & 4.76 & 3.61 & 18 & 384 & 54 & 13.14 && 10 & 100 & 7.74 & 6.52 & 42 & 394 & 94 & 18.45 \\
~4 & ~~1 & 0.22 & 0.20 & 18 & 158 & 32 & ~2.96 && 11 & ~~1 & 0.58 & 0.58 & 38 & 166 & 60 & ~6.38 \\
~4 & ~~3 & 0.43 & 0.38 & 16 & 174 & 36 & ~3.41 && 11 & ~~3 & 1.65 & 1.60 & 36 & 184 & 70 & ~9.50 \\
~4 & ~10 & 1.01 & 0.86 & 18 & 206 & 40 & ~6.41 && 11 & ~10 & 2.92 & 2.75 & 36 & 212 & 70 & 11.11 \\
~4 & ~30 & 2.19 & 1.82 & 18 & 238 & 33 & 10.43 && 11 & ~30 & 4.07 & 3.59 & 36 & 268 & 72 & 11.87 \\
~4 & 100 & 4.60 & 3.66 & 16 & 316 & 54 & 12.10 && 11 & 100 & 7.79 & 6.73 & 36 & 404 & 80 & 15.16 \\
~5 & ~~1 & 0.35 & 0.33 & 18 & 154 & 32 & ~5.16 && 12 & ~~1 & 0.12 & 0.12 & 36 & 150 & 54 & ~2.09 \\
~5 & ~~3 & 0.57 & 0.52 & 18 & 168 & 36 & ~6.12 && 12 & ~~3 & 0.50 & 0.47 & 34 & 170 & 60 & ~5.86 \\
~5 & ~10 & 1.13 & 0.98 & 18 & 200 & 40 & ~8.29 && 12 & ~10 & 1.93 & 1.80 & 36 & 200 & 52 & ~8.89 \\
~5 & ~30 & 2.26 & 1.88 & 18 & 296 & 44 & 10.69 && 12 & ~30 & 4.05 & 3.62 & 34 & 260 & 72 & 11.05 \\
~5 & 100 & 4.34 & 3.37 & 18 & 310 & 52 & 11.18 && 12 & 100 & 7.26 & 6.28 & 36 & 370 & 78 & 16.25 \\
~6 & ~~1 & 0.26 & 0.22 & 18 & 148 & 34 & ~4.00 && 13 & ~~1 & 0.78 & 0.78 & 48 & 168 & 86 & ~7.60 \\
~6 & ~~3 & 0.39 & 0.34 & 18 & 168 & 34 & ~3.74 && 13 & ~~3 & 1.88 & 1.85 & 46 & 182 & 92 & 13.80 \\
~6 & ~10 & 1.14 & 0.99 & 18 & 176 & 40 & ~9.54 && 13 & ~10 & 4.07 & 3.90 & 40 & 198 & 92 & 25.00 \\
~6 & ~30 & 2.25 & 1.87 & 20 & 274 & 44 & 12.69 && 13 & ~30 & 5.53 & 5.00 & 38 & 266 & 94 & 19.78 \\
~6 & 100 & 4.11 & 3.08 & 18 & 342 & 54 & ~9.45 && 13 & 100 & 7.55 & 5.73 & 44 & 378 & 94 & 15.72 \\
~7 & ~~1 & 0.26 & 0.24 & 18 & 158 & 34 & ~3.69 && 14 & ~~1 & 0.12 & 0.11 & 48 & 110 & 52 & ~2.24 \\
~7 & ~~3 & 0.42 & 0.40 & 16 & 170 & 36 & ~3.18 && 14 & ~~3 & 0.55 & 0.52 & 48 & 168 & 88 & ~4.28 \\
~7 & ~10 & 1.15 & 1.05 & 14 & 188 & 46 & ~5.71 && 14 & ~10 & 2.55 & 2.45 & 46 & 192 & 94 & 17.64 \\
~7 & ~30 & 1.86 & 1.67 & 14 & 238 & 40 & ~7.39 && 14 & ~30 & 4.40 & 3.85 & 42 & 232 & 94 & 13.09 \\
~7 & 100 & 4.65 & 4.02 & 16 & 312 & 46 & 15.35 && 14 & 100 & 7.74 & 6.52 & 42 & 394 & 94 & 18.45 \\
\enddata
\tablenotetext{a}{
For each model number, the columns list the initial mass in solids $M_0$, 
the final mass in solids $M_f$, and the final mass in the disk, $M_d$, 
in units of $10^{21}$~g; the inner and outer disk radius, $a_1$ and $a_2$ 
(in units of \rp); the radius of maximum surface density, $a_m$, in \rp, 
and the maximum surface density $\Sigma_m$ in units of \gcms.
}
\end{deluxetable}

\begin{figure}[t]
\begin{center}
\includegraphics[width=5.5in]{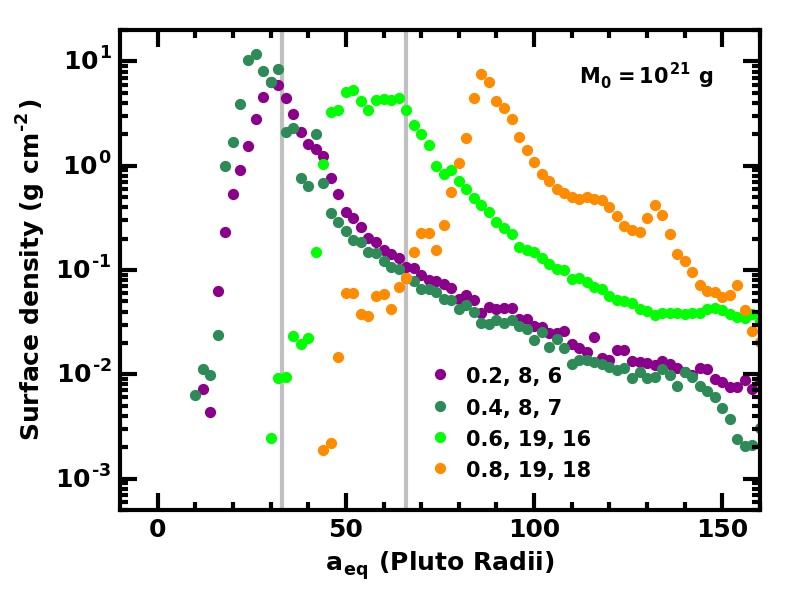}
\vskip -2ex
\caption{\label{fig: sigma4}
Comparison of surface density distributions at 1000~yr for calculations with
\M0\ = $10^{21}$~g and the combinations of $e$, $a_{in}$, and $q_0$ indicated
in the legend. The vertical grey lines indicate the boundaries of the satellite
zone.  In compact \pc\ binaries with $a$ = 5~\rp\ and $e$ = 0.2--0.4,
the surface density reaches a maximum at the inner edge of the satellite zone,
$\sim$ 30--35~\rp. In wider binaries with $a$ = 10~\rp\ and $e$ = 0.6--0.8,
peak surface density lies at larger \aeq, 50--60~\rp when $e$ = 0.6 and 
80--90~\rp when $e$ = 0.8. Systems with $e$ = 0.8 have peak $\Sigma$ outside 
the satellite zone.
}
\end{center}
\end{figure}

The Table collates several interesting features summarized in the discussions
of models 3--8. The ratio of $M_f$ and $M_d$ to the initial mass $M_0$ declines 
with $M_0$. Systems with a small initial mass, $M_0 = 10^{21}$~g, place 10\% to 
30\% of small solids into a disk.  When the initial mass is 100 times larger, 
only $\sim$ 5\% of the initial mass remains in the system. In all cases, the 
inner edge of the disk is 3--4 times the separation of the inner binary, with 
$a_1 \approx$ 15--20~\rp\ when $a$ = 5~\rp\ and $a_1 \approx$ 35--45~\rp when 
$a \approx$ 10~\rp. Peak surface density occurs at roughly twice $a_1$. Although 
$a_1$ and $a_m$ are independent of $M_0$, the outer edge of the disk correlates 
well with $M_0$, ranging from $a_2 \approx$ 125--150~\rp\ for $M_0 = 10^{21}$~g 
to $a_2 \approx$ 300--400~\rp\ for $M_0 = 10^{23}$~g. Systems with large disk 
radii have significant amount of solids outside the satellite zone.

To highlight one outcome from Table~\ref{tab: results}, Fig.~\ref{fig: sigma4}
summarizes differences in the final surface density distributions with different 
central binaries. When the \pc\ binary is compact and nearly circular, peak 
surface density is close to the inner edge of the satellite zone. As $e$ grows,
peak $\Sigma$ moves outward. In \pc\ binaries with $(a, e)$ = (5~\rp, 0.6), the
surface density reaches a broad maximum at 50--60~\pc, roughly coincident with
the orbits of Kerberos and Hydra. For $e$ = 0.8, the edge of the disk rests in
the middle of the satellite zone; $\Sigma$ has a sharp peak at $\sim$ 85~\pc, 
well outside the satellite zone.

In each calculation at $t$ = 1000~yr, very little of the disk lies within regions
where circumbinary orbits are unstable \citep[e.g.,][]{holman1999,doolin2011}.
In numerical simulations of circumbinary test particles, the innermost stable
orbits for systems with mass ratios similar to \pc\ lie at 14~\rp\ ($a, e$ = 
5~\rp, 0.2), 17--18~\rp\ ($a, e$ = 5~\rp, 0.4), 38--42~\rp\ ($a, e$ = 10~\rp, 0.6), 
and 46~\rp\ ($a, e$ = 10~\rp, 0.8). Most of the results in Table~\ref{tab: results}
place the inner edge of the disk in the stable region of the system (see also
Table ~\ref{tab: init}). In the few examples where disk material lies within an 
unstable region, the binary will eject these solids within another few thousand 
years \citep{kb2019a}.

Although the binary clearly sets the inner edge of the disk, collisional damping 
and the gravitational potential of the binary combine with the initial conditions 
to establish the overall disk morphology. In compact binaries with $a$ = 5~\rp, 
some circumbinary solids initially have orbits with semimajor axes that are stable 
for particles on circular orbits. Others have semimajor axes within the unstable 
zone.  As collisional damping circularizes the solids, the binary potential encourages 
them to move onto orbits with larger $a$. Over 1000~yr, the binary evacuates the 
volume where orbits are unstable, leaving behind solids on stable orbits. Because 
solids were nearly on stable orbits at $t$ = 0, the final orbits for many solids lie 
close to the inner edge of the stable region.

In contrast, solids orbiting less compact, more eccentric binaries all initially
lie within the unstable region. Although collisional damping has a somewhat bigger
challenge in circularizing the orbits of these solids, wider binaries with longer 
orbital periods push on the solids more gently. As the solids move farther and
farther away from the binary, damping has time to circularize the orbits. Because
the unstable region around the binary is large, solids have to find stable orbits 
well outside the binary.

\begin{figure}[t]
\begin{center}
\includegraphics[width=5.5in]{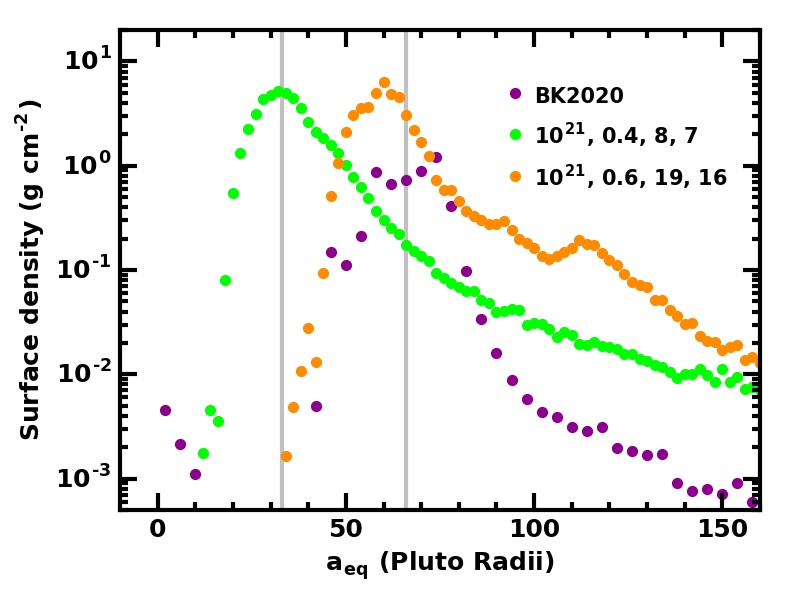}
\vskip -2ex
\caption{\label{fig: sigma5}
Comparison of surface density distributions generated from the impact of a TNO on
Charon \citep[BK2020, purple symbols][]{bk2020} with the final states of debris 
surrounding the giant impact that forms the \pc\ binary. Green (orange) symbols
indicate results for $\M0, e, a_{in}, q_0$ = $10^{21}~{\rm g}, 0.4, 8~\rp, 7~\rp$
($10^{21}~{\rm g}, 0.6, 19~\rp, 16~\rp$). The vertical grey lines indicate the 
boundaries of the satellite zone.
}
\end{center}
\end{figure}

In these examples, particles on circular orbits at the inner edge of the satellite 
zone are stable (unstable) in binaries with $a$ = 5~\rp\ and $e$ = 0.2--0.4 
($a$ = 10~\rp\ and $e$ = 0.6--0.8). This difference results in very different
disk morphologies: circumbinary disks in compact binaries have significant mass 
within the satellite zone; disks in wider binaries have most of their mass outside
the satellite zone.

To conclude this section, Fig.~\ref{fig: sigma5} compares results for $\Sigma$
from model 5 (green symbols, \M0\ = $10^{21}$~g) and model 11 (orange symbols, 
\M0\ = $10^{21}$~g) with the surface density distribution for debris from the
impact of a TNO with Charon \citep[purple symbols;][]{bk2020}. In the TNO collision, 
60\% to 70\% of the $10^{20}$~g ejected by the impact survives to form a circumbinary 
ring.  Although this ring contains less material than required to build the known 
circumbinary satellites, the surface density maximum coincides with the maximum
of disks surrounding a \pc\ binary with $e$ = 0.6. The inner edge of this ring
also roughly overlays the inner edge of the disk around the more eccentric binary.

The sharp outer edge of rings produced from TNO collisions with Charon provides
a sharp contrast with the disks generated from giant impacts. From the peak at
80~\rp\ to 100~\rp, the surface density within the ring drops by nearly three 
orders of magnitude.  Within disks from the giant impact, the decline in $\Sigma$
is a more modest factor of ten. The gradual decline in $\Sigma$ is a hallmark of 
all of the extended disks in our calculations.

\vskip 6ex
\section{DISCUSSION}
\label{sec: disc}

Over the past decade, the giant impact hypothesis has become the most popular
model for the formation of the \pc\ binary planet \citep[e.g.,][and references
therein]{mckinnon2017,stern2018,mckinnon2020}. 
In the commonly accepted picture, two planets orbiting the Sun within 
the protosolar nebula suffer a glancing collision \citep{canup2005,canup2011}.
The lower mass secondary loses momentum and is captured by the more massive 
primary; debris from the collision orbits the newly-formed, compact binary.  
In a more recent option, the planets first form a very wide binary 
\citep{rozner2020}.  Over time, the orbit decays, which facilitates 
a collision with a geometry similar to that in a giant impact.

Immediately after the collision, the circumbinary debris lies on very eccentric orbits
that are intrinsically unstable.  Within this debris, solid particles follow one of 
two dynamical paths \citep{kb2014b,walsh2015,kb2019c,bk2020}.  Solids that do not 
physically collide with other solids cannot circularize their orbits. Within a month, 
the binary either ejects these solids or accretes them.  When an orbiting solid collides 
with another solid, the large relative collision velocities guarantee catastrophic 
disruption; particles within the debris are then much smaller than the original solids. 
After only a few such collisions, 1--10~km solids can be ground down into 1--10~m solids 
\citep{kb2014b,bk2020}. Because small solids damp effectively, this process allows solids 
to circularize their orbits on short time scales. 

In a first attempt to chart the evolution of solids following the giant impact,
\citet{walsh2015} performed a set of numerical calculations of high eccentricity 
particles with initial semimajor axes between the 3:1 and 5:1 orbital resonances 
with Charon\footnote{For a binary semimajor axis $a$ = 10~\rp, the 3:1, 5:1, and 7:1 
resonances have semimajor axes of $\sim$ 20~\rp, 30~\rp, and 36~\rp.}.
Although some solids eventually cross Charon's orbit and are ejected from the binary,
initial orbits have pericenters well outside the apocenter of Charon's orbit.  
Within this suite of calculations, systems with sufficiently large numbers of 
inter-particle collisions per orbit develop circumbinary disks with relatively few 
losses from dynamical ejections. The combination of collisional damping and dynamical 
interactions with Charon generate extended disks on stable circumbinary orbits.  For 
binaries with $e$ = 0.0--0.3, disk material orbits outside the 4:1 orbital resonance 
with Charon and inside the 7:1 resonance. Disks orbiting more eccentric binaries are 
more extended than those orbiting more circular binaries. With little loss in mass from
dynamical ejections, all disks have enough mass in the satellite zone to grow several
10--20~km moons. The lack of mass outside the orbit of Hydra eliminates the possibility
of satellite growth outside the satellite zone.

The calculations described here complement and extend these results. In models 1--14,
the solids begin on eccentric orbits that cross Charon's orbit (models 2, 4, 6, 8, 10, 
12, and 14) or have pericenters at the apocenter of Charon's orbit (models 1, 3, 5, 7,
9, 11, and 13). Compared to \citet{walsh2015}, the central binary has a broader range 
of eccentricity, $e$ = 0.2--0.8 instead of $e$ = 0.0--0.3. The range of initial solid
mass considered here is also somewhat larger than in \citet{walsh2015}.

Despite differences in initial conditions, outcomes are similar. In all of the 70 
calculations discussed here, the binary ejects a substantial fraction of solids from 
the system. Compared to \citet{walsh2015}, the initial orbital states of the solids are 
much more extreme; thus, collisional damping is unable to circularize the orbit of every
solid in the system. Despite the ejections, disk formation is ubiquitous.  The inner
structure of the disk is fairly similar to results described in \citet{walsh2015}, with an 
inner edge just outside the region where circumbinary orbits are unstable and a maximum 
$\Sigma$ at somewhat larger radii. 

However, the disks considered here are much more extensive than those in \citet{walsh2015}, 
with outer radii of 150--400~\rp. The initial conditions are responsible for this difference.
By starting on orbits that cross or nearly cross the orbit of Charon, the solids considered
here receive much larger kicks from the central binary and are pushed much farther away from
the system barycenter. Collisional damping still effectively circularizes orbits, but generates
much larger disks.  Within these disks, the mass in solids ranges from $\sim$ 1 to 50--70 
times the mass in the known \pc\ satellites. 

Although every disk has mass sufficient to form moons like the current \pc\ satellites,
all disks may not generate satellites in their current locations. In systems where \pc\ has
an initial $e$ = 0.2--0.4, the surface density peak lies at the inner edge of the satellite
zone (Fig.~\ref{fig: sigma4}). These disks have a factor of 5--6 less mass near the orbit
of Hydra than near the orbits of Styx and Nix. In contrast, disks orbiting \pc\ binaries 
with $e$ = 0.6 have most of their mass near the orbits of Kerberos and Hydra and relatively
little near Styx--Nix. Finally, systems with an initial $e$ = 0.8 form disks with nearly 
all of their mass outside the satellite zone at $a \gtrsim$ 80~\rp.  

Clearly, understanding which disks form satellites with masses and orbital architectures 
similar to the \pc\ satellites requires a more extensive set of calculations, However,
results in \citet{kb2014b} and \citet{walsh2015} provide useful guidance. Among the 
systems considered here, those with $e \approx$ 0.4--0.6 seem most likely to generate 
large moons in the satellite zone. In binaries with $e$ = 0.2 (0.8), newly-formed moons
would need to migrate outward (inward) to reach the same locations as the known satellites.
For all binaries, disks have a fraction of their mass outside the satellite zone. 
Formation of massive satellites in this region is plausible \citep{kb2014b}. 
If these disks form satellites, the moons must either be small enough to avoid detection 
by \nh\ \citep{weaver2016} or migrate from large distances into the satellite zone 
\citep{kb2014b}. 

For any circumbinary disk in the \pc\ system, several evolutionary processes impact
satellites as they grow within the disk. Right after the giant impact, tidal forces 
begin to circularize and then to expand the \pc\ orbit \citep[e.g.,][]{farinella1979,
dobro1997,peale1999,cheng2014a,correia2020}. If satellites grow rapidly as in 
\citet{walsh2015}, orbital resonances sweep through the satellite zone 
\citep{ward2006,lith2008a,cheng2014b}. Satellite orbits are then rapidly 
destabilized. More slowly growing satellites as in \citet{kb2014b} may retain 
sufficient mass in small particles to stabilize the orbits of large moons
\citep{kb2014b,bk2015b}. Dynamical calculations of massless tracer particles suggest 
a $\sim$ 0.1--10~Myr time scale to remove small particles orbiting between the four 
satellites \citep{kb2014b,kb2019a}. Because tides expand the orbit on a similar time 
scale, it is important to include tidal expansion in calculations of particle growth. 
We plan to complete these types of calculations in the near future.

Aside from tidal evolution of the central binary, gravitational interactions with 
the Sun can influence the formation and evolution of moons in the \pc\ binary 
\citep{michaely2017}.  Recent analyses suggest the Sun truncates any circumbinary 
\pc\ disk at 400--500~\rp.  Although gravity of the Sun does not affect solids 
within the satellite zone, Lidov--Kozai oscillations can place particles in the 
outer disk on high $e$ orbits. These particles might then collide with solids 
in the inner disk, adding their mass to the satellite zone and enhancing the 
formation of large moons. We have not included the Sun in the calculations 
discussed here, but it is straightforward to do so. Adding the Sun might 
reduce the extent of the circumbinary disks and enable the formation of 
more massive satellites in the inner disk.

Despite the popularity of the giant impact picture for satellite formation in \pc, it is
plausible that satellites form in the debris from an impact between Charon and a TNO
\citep{bk2020}. Because debris from such a collision naturally forms a circumbinary ring,
this model is attractive. Satellites form only near or within the satellite zone, with
a much smaller chance of moon formation outside the satellite zone. If the TNO collision
occurs after the completion of tidal evolution, newly-formed satellites need not dodge 
expanding orbital resonances from the binary. The circumbinary rings of \citet{bk2020}
are also safe from Lidov--Kozai oscillations.

Choosing among possible models requires calculations that include the physical processes
outlined above. It seems plausible that at least one configuration of a circumbinary
disk or ring will yield satellites similar to the known satellites. Aside from explaining
the origin of the satellites, these calculations may constrain the properties of the
giant impact (by establishing a preference for a small range of initial binary $a$ and $e$)
or the space density of TNOs after \pc\ reaches its current tidally locked state.

\vskip 6ex
\section{SUMMARY}
\label{sec: summ}

We consider the dynamical evolution of circumbinary solids following the giant impact
that formed \pc. Starting from initial orbits that cross or nearly cross Charon's orbit,
particles respond to the time-varying binary potential and damping from particle-particle
collisions. On time scales of 100--1000~yr, the central binary ejects a large fraction
of the solids into the Solar System. At the same time, collisional damping circularizes 
the orbits of many other particles. The combined action of damping and gravitational 
pushes 
from \pc\ yields a vertically thin circumbinary disk \citep[see also][]{walsh2015}. The 
disk extends from close to the innermost stable circumbinary orbit at 20--40~\rp\ to 
200--400~\rp. Within the disk, the maximum surface density is often within the 
`satellite zone,' the circumbinary volume that contains the orbits of the four 
small satellites orbiting \pc. 

Although all initial conditions lead to disk formation, outcomes depend on the properties
of the \pc\ binary and the initial orbital architecture of the circumbinary debris from 
the giant impact.  Wider binaries with $a$ = 10~\rp\ and $e$ = 0.6--0.8 retain a larger 
fraction of the impact debris than more compact binaries with $a$ = 5~\rp\ and $e$ = 0.2--0.4. 
Debris fields with less mass, $\sim 10^{21}$~g, and a smaller fraction of orbits with 
pericenters inside Charon's orbit preserve more of their initial mass than those with more
mass and Charon-crossing orbits. Despite these differences, all calculations retain enough 
mass to form several 5--20~km satellites. However, some configurations may enable the growth
of large satellites well outside the orbit of Hydra, which is precluded by \nh\ imaging data.

With the calculations of \citet{bk2020}, there are now two plausible paths for the
formation of the \pc\ circumbinary satellites. In the giant impact model, the conditions
within circumbinary disks are favorable for growing small moons in less than 1~Myr. As
these moons form, they must survive tidal expansion of the central binary. In a small 
impact model, debris from the collision between Charon and a TNO evolves into a circumbinary
ring of material roughly coincident with the satellite zone. As the binary has already 
evolved into its current configuration, satellites generated within this ring need not
worry about tidal expansion.

Choosing between these two scenarios requires additional calculations that include tidal
evolution and interaction with the Sun and allow for orbital migration of the satellites.
If a robust choice is possible, these calculations could provide additional constraints 
on the formation of \pc\ by isolating a range of initial $a$ and $e$ that is more conducive
to the growth and survival of Styx, Nix, Kerberos, and Hydra. Outcomes of impacts between 
Charon and large TNOs could yield better estimates of the cratering frequency on \pc\ and
the early evolution of large solids in the Kuiper belt \citep[e.g.,][]{singer2019,kb2020}.

\vskip 6ex

Some of the data (ascii or binary output files and C programs capable of reading them) 
generated from previous numerical studies of the \pc\ system are available at a 
publicly-accessible repository
(https://hive.utah.edu/) with these urls
https://doi.org/10.7278/s50d-w273-1gg0 \citep{kb2019a},
https://doi.org/10.7278/S50D-HAJT-E0G0 \citep{kb2019b},
https://doi.org/10.7278/S50D-EFCY-ZC00 \citep{kb2019c},
https://doi.org/10.7278/S50D4AKFQZFC \citep{bk2020}, and
https://doi.org/10.7278/S50D5Q2MFDBT \citep{kb2020}.
For this paper, data are available at the same repository, with the url
https://doi.org/10.7278/S50DSSMBHHXN.

\vskip 6ex

We acknowledge generous allotments of computer time on the NASA `discover' cluster.
Advice and comments from M. Geller and an anonymous referee improved the presentation
of these results.
Portions of this project were supported by the {\it NASA } {\it Outer Planets} and 
{\it Emerging Worlds} programs through grants NNX11AM37G and NNX17AE24G.

\bibliography{ms.bbl}

\end{document}

%% file: defs.tex
\def\nbody{$n$-body}

\def\deg{\ifmmode {^\circ}\else {$^\circ$}\fi}
\def\degree{\ifmmode {^\circ}\else {$^\circ$}\fi}
\def\mum{\ifmmode {\rm \,\mu {\rm m}}\else $\rm \,\mu {\rm m}$\fi}
\def\arcsec{\ifmmode ^{\prime \prime}\else $^{\prime \prime}$\fi}

\def\inch{\ifmmode ^{\prime \prime}\else $^{\prime \prime}$\fi}
\def\gs{\ifmmode {{\rm g~s^{-1}}}\else ${\rm g~s^{-1}}$\fi}
\def\msunyr{\ifmmode {M_{\odot}~{\rm yr^{-1}}}\else $M_{\odot}~{\rm yr^{-1}}$\fi}
\def\msun{\ifmmode {M_{\odot}}\else $M_{\odot}$\fi}
\def\rsun{\ifmmode {R_{\odot}}\else $R_{\odot}$\fi}
\def\lsun{\ifmmode {L_{\odot}}\else $L_{\odot}$\fi}
\def\mstar{\ifmmode {M_{\star}}\else $M_{\star}$\fi}
\def\rstar{\ifmmode {R_{\star}}\else $R_{\star}$\fi}
\def\tstar{\ifmmode {T_{\star}}\else $T_{\star}$\fi}
\def\lstar{\ifmmode {L_{\star}}\else $L_{\star}$\fi}
\def\mwd{\ifmmode {M_{wd}}\else $M_{wd}$\fi}
\def\rwd{\ifmmode {R_{wd}}\else $R_{wd}$\fi}
\def\twd{\ifmmode {T_{wd}}\else $T_{wd}$\fi}
\def\lwd{\ifmmode {L_{wd}}\else $L_{wd}$\fi}
\def\md{\ifmmode {M_d}\else $M_d$\fi}
\def\ld{\ifmmode {L_d}\else $L_d$\fi}
\def\ad{\ifmmode A_d\else $A_d$\fi}
\def\ldlwd{\ifmmode L_d / L_{wd}\else $L_d / L_{wd}$\fi}
\def\ldlstar{\ifmmode L_d / L_\star\else $L_d / L_{\star}$\fi}
\def\rearth{\ifmmode {\rm R_{\oplus}}\else $\rm R_{\oplus}$\fi}
\def\mearth{\ifmmode {\rm M_{\oplus}}\else $\rm M_{\oplus}$\fi}
\def\qc{\ifmmode Q_c\else $Q_c$\fi}
\def\qdstar{\ifmmode Q_D^\star\else $Q_D^\star$\fi}
\def\rt{\ifmmode r_t\else $r_t$\fi}
\def\vc{\ifmmode v_c\else $v_c$\fi}
\def\vsqd{\ifmmode v^2 / Q_D^\star\else $v^2 / Q_D^\star$\fi}
\def\kms{\ifmmode {\rm km~s^{-1}}\else $\rm km~s^{-1}$\fi}
\def\ms{\ifmmode {\rm m~s^{-1}}\else $\rm m~s^{-1}$\fi}
\def\vrel{\ifmmode v_{rel}\else $v_{rel}$\fi}
\def\mdot{\ifmmode \dot{M}\else $\dot{M}$\fi}
\def\mdotz{\ifmmode \dot{M}_0\else $\dot{M}_0$\fi}
\def\mesc{\ifmmode m_{esc}\else $m_{esc}$\fi}
\def\aeq{\ifmmode a_{eq}\else $a_{eq}$\fi}
\def\rmin{\ifmmode r_{min}\else $r_{min}$\fi}
\def\rmax{\ifmmode r_{max}\else $r_{max}$\fi}
\def\xmax{\ifmmode x_{max}\else $x_{max}$\fi}
\def\mmin{\ifmmode m_{min}\else $m_{min}$\fi}
\def\mmax{\ifmmode m_{max}\else $m_{max}$\fi}
\def\rmind{\ifmmode r_{min,d}\else $r_{min,d}$\fi}
\def\rmaxd{\ifmmode r_{max,d}\else $r_{max,d}$\fi}
\def\mmaxd{\ifmmode m_{max,d}\else $m_{max,d}$\fi}
\def\vrad{\ifmmode v_{rad}\else $v_{rad}$\fi}
\def\qz{\ifmmode q_{0}\else $q_{0}$\fi}
\def\qi{\ifmmode q_{i}\else $q_{i}$\fi}
\def\ql{\ifmmode q_{l}\else $q_{l}$\fi}
\def\qs{\ifmmode q_{s}\else $q_{s}$\fi}
\def\vhill{\ifmmode v_H\else $r_H$\fi}
\def\rhill{\ifmmode r_H\else $r_H$\fi}
\def\Rhill{\ifmmode R_H\else $R_H$\fi}
\def\rbrk{\ifmmode r_{brk}\else $r_{brk}$\fi}
\def\rdamp{\ifmmode r_{damp}\else $r_{damp}$\fi}
\def\rin{\ifmmode r_{in}\else $r_{in}$\fi}
\def\rout{\ifmmode r_{out}\else $r_{out}$\fi}
\def\tin{\ifmmode t_{in}\else $t_{in}$\fi}
\def\tout{\ifmmode t_{out}\else $t_{out}$\fi}
\def\ain{\ifmmode a_{in}\else $a_{in}$\fi}
\def\aout{\ifmmode a_{out}\else $a_{out}$\fi}
\def\r0{\ifmmode r_{0}\else $r_{0}$\fi}
\def\R0{\ifmmode R_{0}\else $R_{0}$\fi}
\def\m0{\ifmmode m_{0}\else $m_{0}$\fi}
\def\mone{\ifmmode m_{1}\else $m_{1}$\fi}
\def\mtwo{\ifmmode m_{2}\else $m_{2}$\fi}
\def\atwo{\ifmmode a_{2}\else $a_{2}$\fi}
\def\etwo{\ifmmode e_{2}\else $e_{2}$\fi}
\def\mf{\ifmmode m_{f}\else $m_{f}$\fi}
\def\af{\ifmmode a_{f}\else $a_{f}$\fi}
\def\ef{\ifmmode e_{f}\else $e_{f}$\fi}
\def\M0{\ifmmode M_{0}\else $M_{0}$\fi}
\def\amax{\ifmmode a_{max}\else $a_{max}$\fi}
\def\a0{\ifmmode a_{0}\else $a_{0}$\fi}
\def\e0{\ifmmode e_{0}\else $e_{0}$\fi}
\def\v0{\ifmmode v_{0}\else $v_{0}$\fi}
\def\xm{\ifmmode x_{m}\else $x_{m}$\fi}
\def\sigz{\ifmmode \Sigma_0\else $\Sigma_0$\fi}
\def\ergg{\ifmmode {\rm erg~g^{-1}}\else ${\rm erg~g^{-1}}$\fi}
\def\ergs{\ifmmode {\rm erg~s^{-1}}\else ${\rm erg~s^{-1}}$\fi}
\def\gyr{\ifmmode {\rm g~yr^{-1}}\else ${\rm g~yr^{-1}}$\fi}
\def\cms{\ifmmode {\rm cm~s^{-1}}\else ${\rm cm~s^{-1}}$\fi}
\def\gcms{\ifmmode {\rm g~cm^{-2}}\else $\rm g~cm^{-2}$\fi}
\def\gcmc{\ifmmode {\rm g~cm^{-3}}\else $\rm g~cm^{-3}$\fi}
\def\atil{\ifmmode {\tilde{a}}\else $\tilde{a}$\fi}
\def\ttil{\ifmmode {\tilde{t}}\else $\tilde{t}$\fi}
\def\sqrttt{\ifmmode {\tilde{t}^{1/2}}\else $\tilde{t}^{1/2}$\fi}

\def\orch{{\it Orchestra}}
\def\nh{{\it New Horizons}}
\def\pc{Pluto--Charon}
\def\mp{\ifmmode m_P\else $m_P$\fi}
\def\mc{\ifmmode m_C\else $m_C$\fi}
\def\mh{\ifmmode m_H\else $m_H$\fi}
\def\mk{\ifmmode m_K\else $m_K$\fi}
\def\ms{\ifmmode m_S\else $m_S$\fi}
\def\mn{\ifmmode m_N\else $m_N$\fi}
\def\rp{\ifmmode R_P\else $R_P$\fi}
\def\rc{\ifmmode R_C\else $R_C$\fi}
\def\apc{\ifmmode a_{PC}\else $a_{PC}$\fi}
\def\mpc{\ifmmode m_{PC}\else $m_{PC}$\fi}
\def\epc{\ifmmode e_{PC}\else $e_{PC}$\fi}